\documentclass[aps,prd,twocolumn,twoside,superscriptaddress,floatfix, nofootinbib]{revtex4-1}
\pdfoutput=1
\usepackage{ifthen,amsmath,amssymb, bm}
\usepackage{graphicx}
\usepackage{hyperref}
\usepackage{float}
\usepackage{aas_macros}
\usepackage{caption}
\usepackage{subcaption}
\usepackage[export]{adjustbox}
\usepackage{bookmark}

\usepackage{newtxtext,newtxmath}
\usepackage{amsmath}
\usepackage{amssymb}
\usepackage{graphicx}
\usepackage{dcolumn}
\usepackage{bm}
\usepackage[dvipsnames,table]{xcolor}

\newcommand{\beq} {\begin{equation}}
\newcommand{\eeq} {\end{equation}}
\newcommand{\bal} {\begin{aligned}}
\newcommand{\eal} {\end{aligned}}

\def\hk{{\hat k}}
\def\hr{{\hat r}}

\newcommand{\boldv}{\boldsymbol v}
\newcommand{\boldr}{\boldsymbol r}
\newcommand{\bx}{\boldsymbol x}
\newcommand{\bk}{\boldsymbol k}
\newcommand{\bvel}{\boldsymbol v}
\newcommand{\bn}{\boldsymbol n}
\newcommand{\bPsi}{\boldsymbol{\Psi}}

\newcommand{\be}{\begin{equation}}
\newcommand{\ee}{\end{equation}}
\newcommand{\ba}{\begin{eqnarray}}
\newcommand{\ea}{\end{eqnarray}}

\def\mpcoh{\,h^{-1}{\rm Mpc}}

\def\msunoh{\,h^{-1}{\rm M}_\odot}

\newcommand{\BH}[1]{\textcolor{blue}{{\bf BH:~{#1}}}}
\newcommand{\SF}[1]{\textcolor{Red}{[SF: #1]}}

\begin{document}

\title{Velocity reconstruction in the era of DESI and Rubin (part II):\\ Realistic samples on the light cone}

\author{Boryana Hadzhiyska}
\email{boryanah@alumni.princeton.edu}
\affiliation{Miller Institute for Basic Research in Science, University of California, Berkeley, CA, 94720, USA}
\affiliation{Physics Division, Lawrence Berkeley National Laboratory, Berkeley, CA 94720, USA}
\affiliation{Berkeley Center for Cosmological Physics, Department of Physics, University of California, Berkeley, CA 94720, USA}
\author{Simone Ferraro}
\affiliation{Physics Division, Lawrence Berkeley National Laboratory, Berkeley, CA 94720, USA}
\affiliation{Berkeley Center for Cosmological Physics, Department of Physics, University of California, Berkeley, CA 94720, USA}
\author{Bernardita Ried Guachalla}
\affiliation{Department of Physics, Stanford University, Stanford, CA, USA 94305-4085}
\affiliation{Kavli Institute for Particle Astrophysics and Cosmology, 382 Via Pueblo Mall Stanford, CA 94305-4060, USA}
\affiliation{SLAC National Accelerator Laboratory 2575 Sand Hill Road Menlo Park, California 94025, USA}
\author{Emmanuel Schaan}
\affiliation{Kavli Institute for Particle Astrophysics and Cosmology,
382 Via Pueblo Mall Stanford, CA 94305-4060, USA}
\affiliation{SLAC National Accelerator Laboratory 2575 Sand Hill Road Menlo Park, California 94025, USA}

\begin{abstract}

Reconstructing the galaxy peculiar velocity field from the distribution of large-scale structure plays an important role in cosmology.
On one hand, it gives us an insight into structure formation and gravity; 
on the other, it allows us to selectively extract the kinetic Sunyaev-Zeldovich (kSZ) effect from cosmic microwave background (CMB) maps. 
In this work, we employ high-accuracy synthetic galaxy catalogs on the light cone to investigate how well we can recover the velocity field when utilizing the three-dimensional spatial distribution of the galaxies in a modern large-scale structure experiment such as the Dark Energy Spectroscopic Instrument (DESI) and the Rubin Observatory Legacy Survey of Space and Time (LSST). 
In particular, we adopt the standard technique used in baryon acoustic oscillation (BAO) analysis for reconstructing the Zeldovich displacements of galaxies through the continuity equation, which yields a first-order approximation to their large-scale velocities. 
We investigate 
variations in the number density, bias, mask, area, redshift noise, and survey depth, as well as modifications to the settings of the standard reconstruction algorithm. 
Since our main goal is to provide guidance for planned kSZ analysis between DESI and the Atacama Cosmology Telescope (ACT), we apply velocity reconstruction to a faithful representation of DESI spectroscopic and photometric targets.
We report the cross-correlation coefficient between the reconstructed and the true velocities along the line of sight. 
For the DESI Y1 spectroscopic survey, we expect the correlation coefficient to be $r \approx 0.64$, while for a photometric survey with $\sigma_z/(1+z) = 0.02$, as is approximately the case for the Legacy Survey used in the target selection of DESI galaxies, $r$ shrinks by half to $r \approx 0.31$.
We hope the results in this paper can be used to inform future kSZ stacking studies and other velocity reconstruction analyses planned with the next generation of cosmology experiments.
All scripts used in this paper can be found here: \url{https://github.com/boryanah/abacus_kSZ_recon}.

\end{abstract}
\maketitle

\section{Introduction}
\label{sec:intro}

The ``missing baryon'' 
problem constitutes one of the most important unresolved mysteries of astrophysics, as it holds the key to understanding the rich physical processes involved in galaxy formation and evolution \cite{2004ApJ...616..643F,2006ApJ...650..560C}. Additionally, baryons account for roughly 15\% of the total matter in the Universe, so having a firm grasp of their distribution is of utmost importance to performing sub-percent precision analysis of upcoming large-scale structure experiments such as the Dark Energy Spectroscopic Instrument \citep[DESI;][]{2016arXiv161100036D,2019BAAS...51g..57L}, the Euclid space telescope \cite{2013LRR....16....6A}, the Rubin Observatory Legacy Survey of Space and Time \citep[LSST;][]{2012arXiv1211.0310L,2019ApJ...873..111I}, and the Nancy Grace Roman Space Telescope \cite{2015arXiv150303757S}. At the crux of this problem is the observation that the baryons residing within galaxies make up only about 10\% of the total cosmological abundance of baryons \cite{2004ApJ...616..643F}. The remaining majority is believed to be located outside of the virial radius of galaxies, in an ionized, diffuse gas, known as the warm-hot intergalactic medium (WHIM). 

Studies of quasar absorption lines \cite{2018Natur.558..406N,2019ApJ...872...83K}, X-ray observations \cite{1998ApJ...495...80B,2002ARA&A..40..539R,2009ApJ...693.1142S}, and Fast Radio Bursts (FRBs) systems \cite{2018PhRvD..98j3518M,2019PhRvD.100j3532M,2020Natur.581..391M} have made some progress in itemizing the baryon content of the Universe. However, the analysis of absorption lines requires the robust modeling of poorly understood astrophysics such as metallicity profiles, whereas X-ray analyses rely on the accurate modeling of gas temperature. As such, these approaches are often limited to a small number of high-mass nearby objects. On the other hand, Sunyaev-Zel’dovich (SZ) measurements provide an alternative path to thoroughly charting the baryons in the Universe. The ionized gas surrounding galaxy clusters leaves distinct imprints (``shadows'') on the cosmic microwave background (CMB) resulting from the interactions between the free electrons in the cluster and the CMB photons. In particular, the two main causes are the bulk motion of the gas, resulting in the kinematic Sunyaev-Zel’dovich (kSZ) effect, and the velocity dispersion of the gas, resulting in the thermal Sunyaev-Zel’dovich effect \citep[tSZ;][]{1972CoASP...4..173S,1980MNRAS.190..413S}. Contrary to other probes, since both SZ effects are largely independent of redshift, they are well-suited for studying clusters throughout cosmic time. While the tSZ signal depends on both the electron number density and the cluster temperature, the kSZ signal is linearly proportional to the electron number density, so it is well-suited to probing low-density regions and the outskirts of low-mass clusters and groups.

However, if we naively stack the kSZ signal around a sample of galaxies or galaxy clusters, the effect would cancel out, as each object has an equal chance of moving towards us or away from us. For this reason, we need to employ an estimate of the peculiar velocity of each galaxy, reconstructed from the three-dimensional galaxy overdensity field. In particular, we can obtain an estimate of the line-of-sight velocity field by solving the linearized continuity equation in redshift space \cite{Padmanabhan12}, similarly to the reconstruction method applied in Baryon Acoustic Oscillations (BAO) analysis. 
This kSZ detection method has been successfully implemented in \cite{Planck16ksz, Hernandez15, Schaan16, Schaan21}, and appears promising for future CMB and large-scale structure experiments \cite{Battaglia17}.
In standard BAO analysis, the goal of reconstruction is to improve the distance-redshift relation \cite{Dodelson20}. The BAO peak is a feature in the 2-point function whose physical size is known. Comparisons of the observed size of the peak with its physical size yield the angular diameter distance and Hubble parameter as a function of redshift. Due to nonlinear evolution, the oscillations on small scales are dampened, which broadens the peak and reduces the accuracy with which its size can be measured. For this reason, all recent BAO surveys apply density-field reconstruction \cite{ESSS07,Padmanabhan12, Zhu:2019gzu}, adopting the first-order (Zel'dovich) approximation, to regain some of the information lost due to the large-scale displacements.

The focus of this paper is to study the accuracy of velocity reconstruction in the controlled setup of synthetic galaxy catalogs, i.e. simulated mocks, created on the light cone. 
The advantage of utilizing light cones is that we can incorporate various realism realistic effects such as redshift-evolution of the galaxy population (e.g., redshift-dependent number density and linear bias), finite area and volume, mask and boundary effects. 
In addition, we estimate the importance of accurate redshift measurements, which is particularly relevant for the analysis of future photometric surveys. 
Employing
the standard method used by the BAO community, i.e., the \texttt{MultiGrid} implementation of \cite{2015MNRAS.450.3822W} via the package `pyrecon'\footnote{\url{https://github.com/cosmodesi/pyrecon}},
we study different tracers and combinations of tracers so as to mimic the high number densities expected to be achieved by future large-scale structure surveys.

In the course of writing this paper, a similar paper came out \cite{2023arXiv231112611C}, interested in using reconstruction in the presence of photometric redshifts for the purposes of BAO science. This is different from the present work, as both the science application and the mock setup are different (we use light cone catalogs rather than cubic mocks).

This paper is organized as follows. Section~\ref{sec:methods} introduces the simulated mocks and the velocity reconstruction method adopted in this study. In Section~\ref{sec:results}, we display the performance of the velocity reconstruction in various scenarios relevant to current and future redshift and photometric surveys. Finally, we summarize our main findings in Section~\ref{sec:conc}.

\section{Methods}
\label{sec:methods}

In this Section, we summarize the simulation products and galaxy mock catalogs used in this study as well as the reconstruction method we adopt to obtain an estimate of the galaxy velocities through the galaxy overdensity field. 

\subsection{\textsc{AbacusSummit} simulation suite}
\label{sec:abacus}

\begin{figure}[H]
    \centering
    \includegraphics[width=0.48\textwidth]{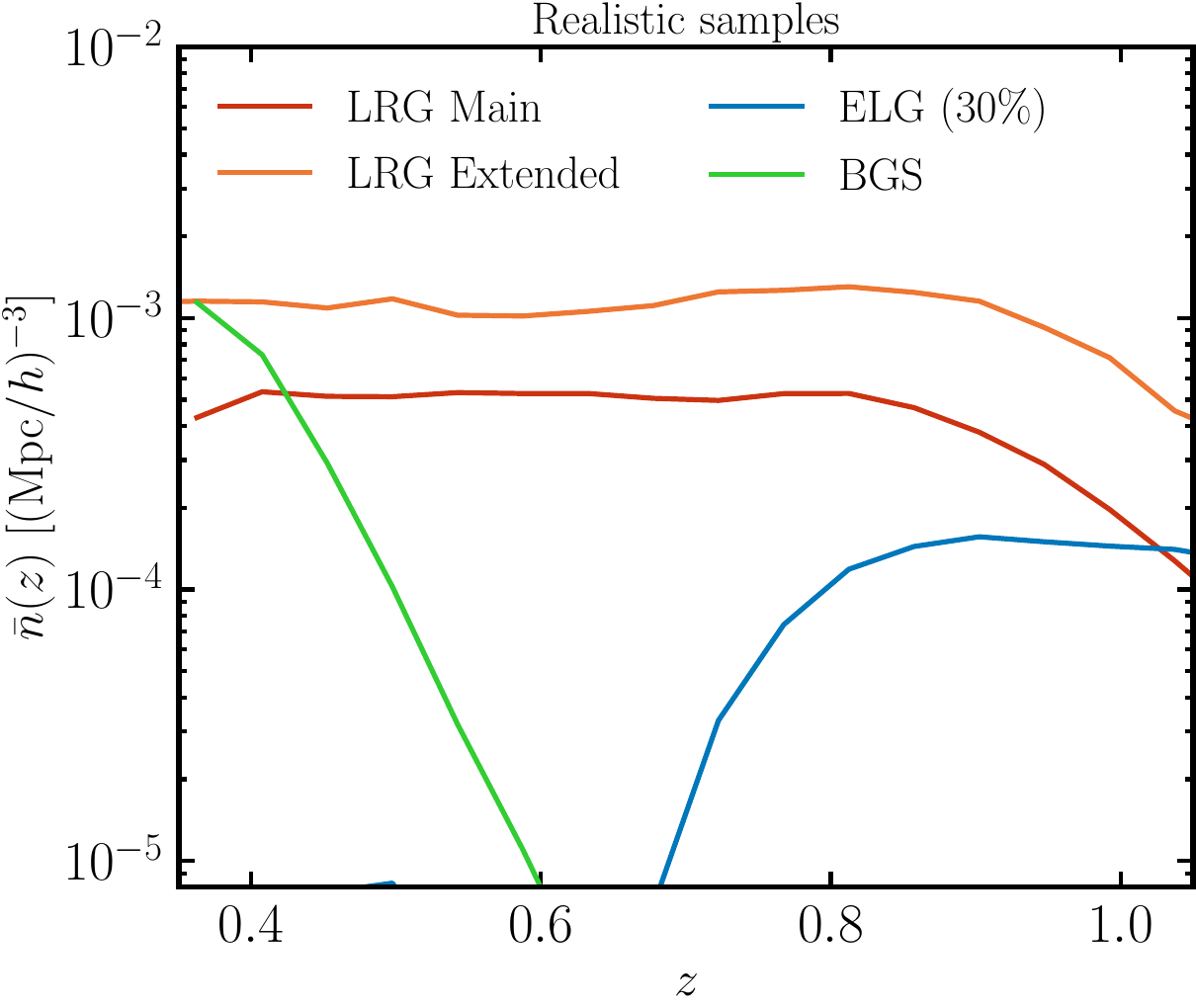}
    \includegraphics[width=0.48\textwidth]{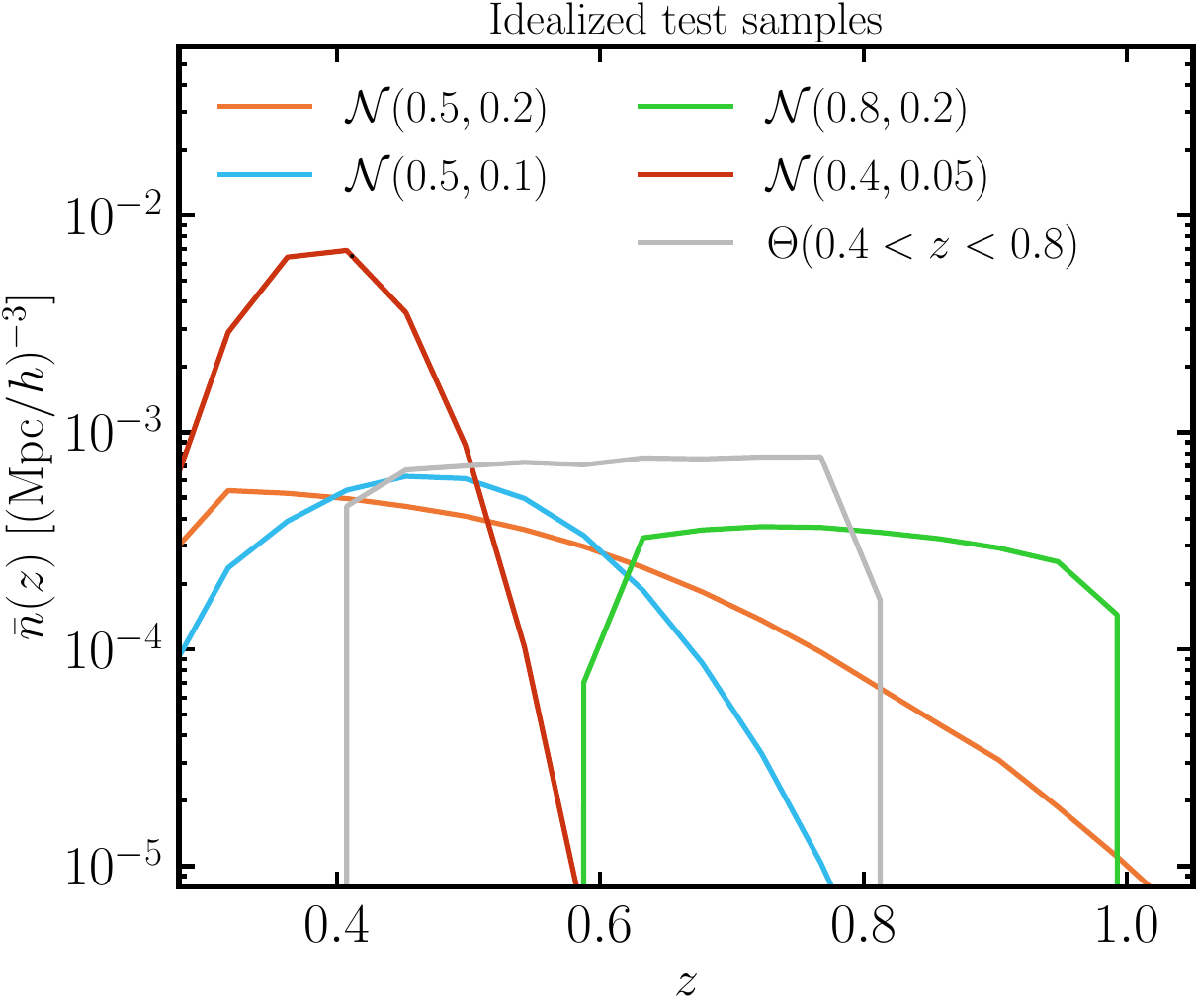}
    \caption{Top panel: Comoving number densities, $\bar n(z)$, as a function of redshift for the four different tracers relevant to this study: Main LRGs, Extended LRGs, ELGs, and BGS. We note that the LRGs appear to have a roughly constant number density of $5.2 \times (\mpcoh)^{-3}$ between $z = 0.4$ and $z = 0.8$; the ELG sample flattens out beyond $z = 0.85$ for the redshifts shown. The BGS objects, on the other hand, have a rapidly declining number density past $z = 0.5$.
    Bottom panel: Comoving number densities, $\bar n(z)$, as a function of redshift for the idealized tests conducted in this study (Section~\ref{sec:ideal}), characterized by simplistic $N(z)$ distributions and no mask effects. In particular, we display the case of Gaussian $N(z)$ functions centered at 0.4, 0.5 and 0.8 as well as a step function between $z = 0.4$ and $0.8$.
    }
    \label{fig:nz}
\end{figure}

In this study we use products from the \textsc{AbacusSummit} suite of high-performance cosmological $N$-body simulations. \textsc{AbacusSummit} \cite{2021MNRAS.508.4017M} was designed to meet and exceed the Cosmological Simulation Requirements of the DESI survey. The simulations were run with \textsc{Abacus} \cite{2019MNRAS.485.3370G,2021MNRAS.508..575G}, a high-accuracy cosmological $N$-body simulation code, optimized for GPU architectures and  for large-volume simulations, on the Summit supercomputer at the Oak Ridge Leadership Computing Facility. 

The majority of the \textsc{AbacusSummit} simulations are made up of the so-called \texttt{base} resolution boxes, which house 6912$^3$ particles in a 2 Gpc$/h$ box, each with a mass of $M_{\rm part} = 2.1 \times 10^9\msunoh$. Additionally, we utilize the \texttt{huge} boxes with corresponding dimensions of 8640$^3$ particles in a 7.5 Gpc$/h$ box (with particle mass of $M_{\rm part} = 5.6 \times 10^{10}\msunoh$). While the \textsc{AbacusSummit} suite spans a wide range of cosmologies, here we focus on the fiducial outputs ($\Omega_b h^2 = 0.02237$, $\Omega_c h^2 = 0.12$, $h = 0.6736$, 
$10^9 A_s = 2.0830$, $n_s = 0.9649$, $w_0 = -1$, $w_a = 0$), consisting of 25 \texttt{base} (\texttt{AbacusSummit\_base\_c000\_ph\{000-024\}}) and two \texttt{huge} runs (\texttt{AbacusSummit\_huge\_c000\_ph\{201,202\}}). For full details on all data products, see \cite{2021MNRAS.508.4017M}.

\subsection{Galaxy samples}
\label{sec:galaxy}

In order to create a maximally realistic catalog of DESI-like objects, we utilize the highly accurate halo light cone catalogs of \textsc{AbacusSummit}, introduced in \cite{2022MNRAS.509.2194H}
\footnote{The \textsc{AbacusSummit} halo light cone catalogues are publicly available at DOI:\href{https://www.doi.org/10.13139/OLCF/1825069}{10.13139/OLCF/1825069}.}. For a discussion of the procedure for generating them and the validation tests performed on them, we refer the reader to \cite{2022MNRAS.509.2194H}. The halo light cone catalogs are designed to efficiently generate mock catalogues via \textsc{AbacusHOD}, a sophisticated routine that builds upon the baseline halo occupation distribution (HOD) model by incorporating various extensions. In this work, we are interested in three different tracer types relevant to DESI: emission-line galaxies (ELGs), luminous red galaxies (LRGs), and the Bright Galaxy Sample (BGS) objects. The \textsc{AbacusHOD} model is described in detail in \cite{2022MNRAS.510.3301Y}.

To obtain the HOD parameters of our DESI-like LRG and ELG samples, we follow the same approach as \cite{2023arXiv230511935H}. 
Namely, we use the best fit HOD values obtained by fitting the correlation function, $\xi(r_p, \pi)$, on small scales ($r \lesssim 30\mpcoh$) from the DESI Survey Validation 3 (SV3) data at two distinct (`pivot') redshifts for each tracer: $z = \{0.5, \ 0.8\}$ for the LRGs and $z = \{0.8, \ 1.1\}$ for the ELGs. The values of the HOD parameters we adopt are shown in Table 1 of \cite{2023arXiv230511935H}. To emulate the intrinsic redshift-dependent change in the galaxy population, we implement a simple redshift-dependent HOD, which linearly interpolates (in scale factor) between the HOD parameter values at the `pivot' redshifts for each tracer. Since the main target of this study is the DESI LRG sample, we focus on the redshift epochs between $z = 0.3$ and $z = 1.1$\footnote{This choice corresponds to the following redshift epochs: $z = 0.3, \ 0.35, \ 0.4, \ 0.45, \ 0.5, \ 0.575, \ 0.65, \ 0.725, \\ \ 0.8, \ 0.875, \ 0.95, \ 1.025, \ 1.1$.}, corresponding to the range for which the LRG number density of the LRGs peaks. The recently published fits of \cite{2023arXiv230606314Y} exhibit nearly matching HOD parameters.

Since the ELG sample obtained in the aforementioned manner has very low bias, and we know that reconstruction is more sensitive to bias than number density, throughout this paper we instead work with a higher-bias ELG sample constructed as follows. We increase the values of the original ELG HOD parameters, $\log M_{\rm cut}$ and $\log M_1$, by $\Delta \log M = 0.5$, then uniformly downsample the thus-obtained catalogs, such that only 30\%  of the galaxies are kept. 
This choice is informed by hydro simulations \citep[for a study of ELGs in hydro simulations, see e.g.,][]{2021MNRAS.502.3599H}, for which we have tested how the HOD parameters change when selecting the 30\% most stellar-massive galaxies. Such a selection is also possible in observations. In that case, the stellar mass is not a direct observable, but we can make a selection on the brightness or luminosity of the galaxies, which is tightly related to their stellar masses. We refer to the thus-constructed `30\% brightest ELG sample' simply as the `ELG sample' throughout the paper.

\begin{figure}[H]
    \centering
    \includegraphics[width=0.48\textwidth]{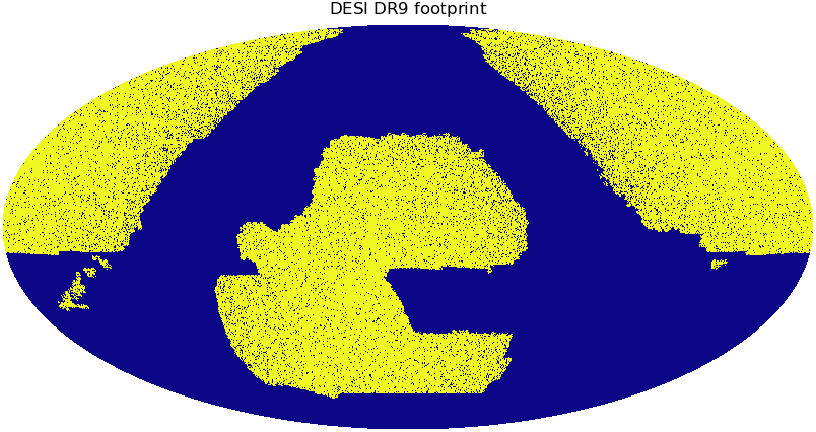}
    \caption{Mollweide projection of the full-sky Data Release 9 (DR9) survey mask for LRGs in equatorial coordinates, constructed by coadding a quality mask (\texttt{BITMASK}) and a number of exposures map (\texttt{GRZMASK}) in order to make sure that the tile has been observed once or more in each magnitude bin ($g$, $r$, $z$) with sufficiently good quality. There are two disjoint regions, corresponding to the North Galactic Cap (NGC) and the South Galactic Cap (SGC).} 
    \label{fig:DESI_mask}
\end{figure}

In addition to the LRGs and ELGs, we add two more galaxy samples: a high-density LRG sample, which aims to mimic the `Extended LRG' catalog \cite{2023AJ....165...58Z}, and a low-mass LRG-like sample, which corresponds to the magnitude-limited BGS. 
We take the best-fit HOD parameter values from fits to the DESI 1\% survey data \citep[the values of the HOD parameters are taken from the first line of Table 4 in][]{2023arXiv230606315P}. To mimic the effect of magnitude limit selection, which preferentially picks low-mass galaxies at low redshifts and high-mass galaxies at higher redshifts, we increase the values of the two HOD parameters controlling the bias, $\log M_{\rm cut}$ and $\log M_1$, by $\Delta \log M = 0.3$ at the higher pivot redshift (remember that we linearly interpolate the HOD parameter values). 
We note that the HOD parameter values at the higher pivot redshift are chosen in an ad-hoc manner, but have a negligible effect on the reconstruction performance, as the BGS number density plummets past $z > 0.5$.
For the Extended LRG sample, we lower the mass threshold parameters, $\log M_{\rm cut}$ and $\log M_1$, at the two pivot redshifts uniformly by $\Delta \log 
M = 0.2$. We check that the bias and the number density of these two new catalogs roughly match the expected values from the published early DESI clustering measurements \cite{2023arXiv230606314Y}. 


\begin{figure}[H]
    \centering
    \includegraphics[width=0.48\textwidth]{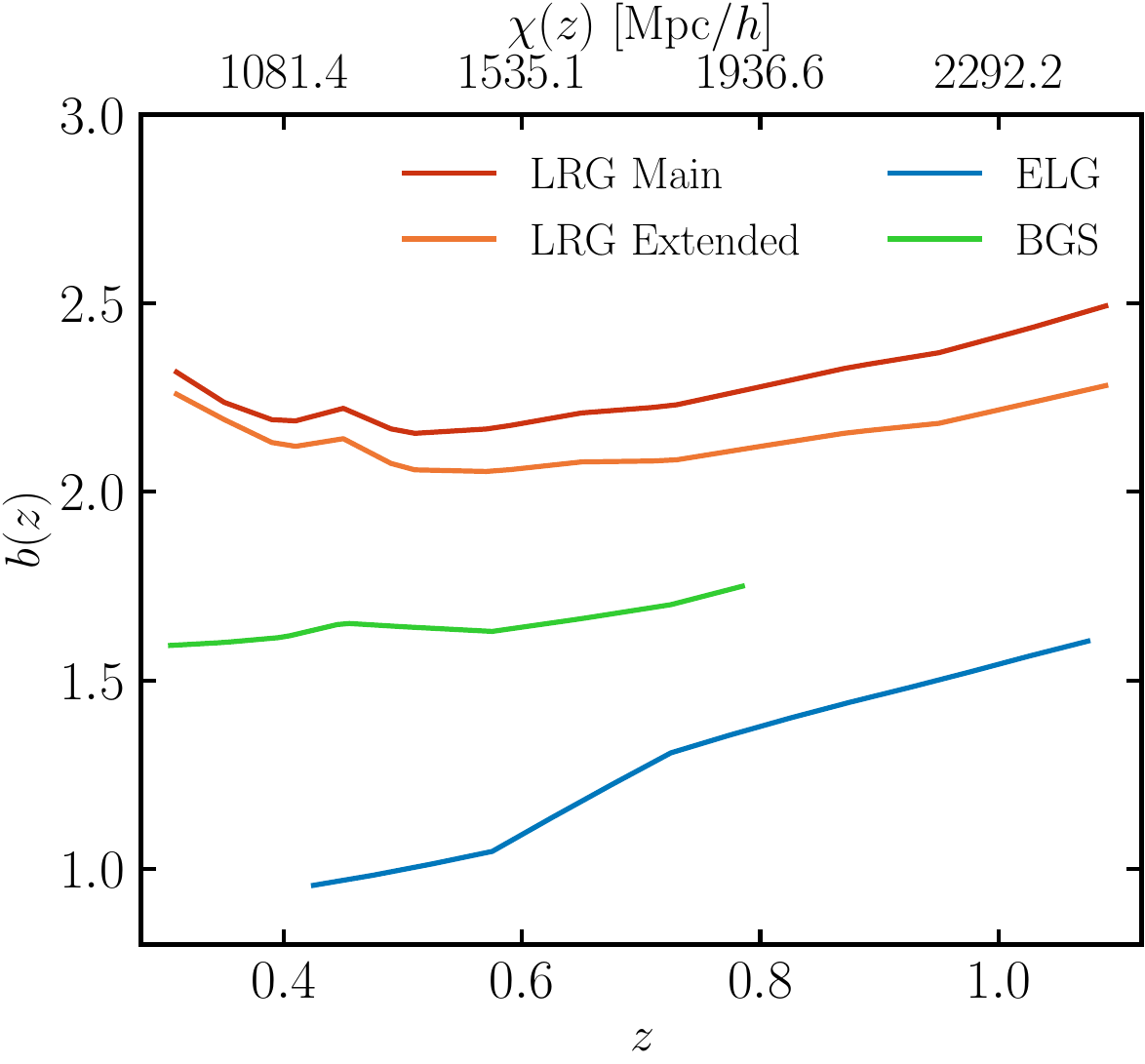}
    \caption{
    Linear bias, $b(z)$, as a function of redshift for the four different tracers relevant to this study: Main LRGs, Extended LRGs, ELGs, and BGS. The bias is computed using the \cite{2010ApJ...724..878T} empirical mass-bias relationship for the host halos. We see that the LRG bias increases slowly for $z \gtrsim 0.4$ due to the evolution of halo bias with redshift. At lower redshifts, $z \lesssim 0.5$, the bias of the LRGs seems to decrease with redshift, which we attribute to the extrapolation below the fitted redshifts. The Extended LRGs have a bias that is lower by $\Delta b = 0.1$ at $z \sim 0.5$. The bias of the BGS and ELGs increases with redshift, as expected, and remains significantly lower than the LRGs.
    }
    \label{fig:bz}
\end{figure}

\begin{figure*}
    \centering
    \includegraphics[width=0.48\textwidth]{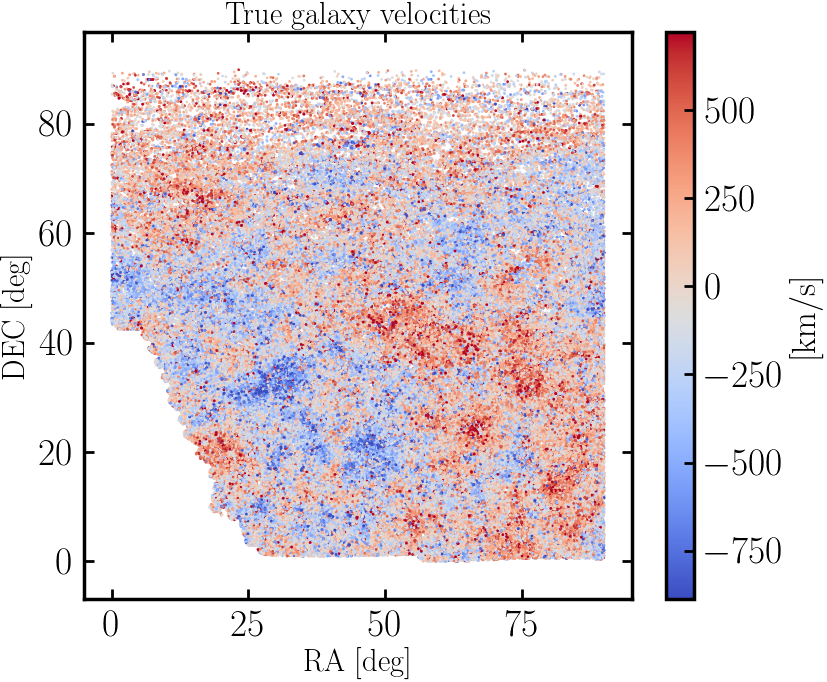}
    \includegraphics[width=0.48\textwidth]{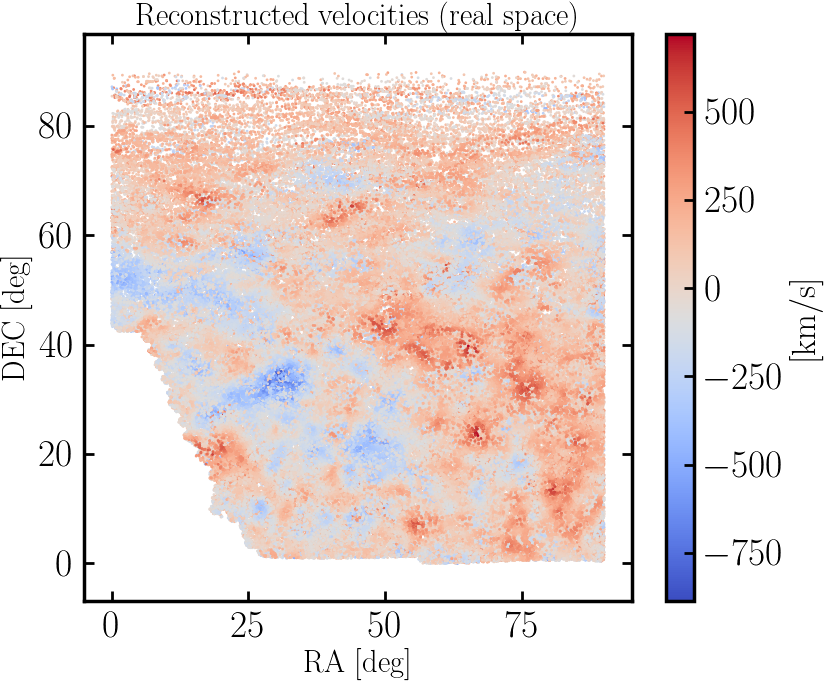}\\
    \includegraphics[width=0.48\textwidth]{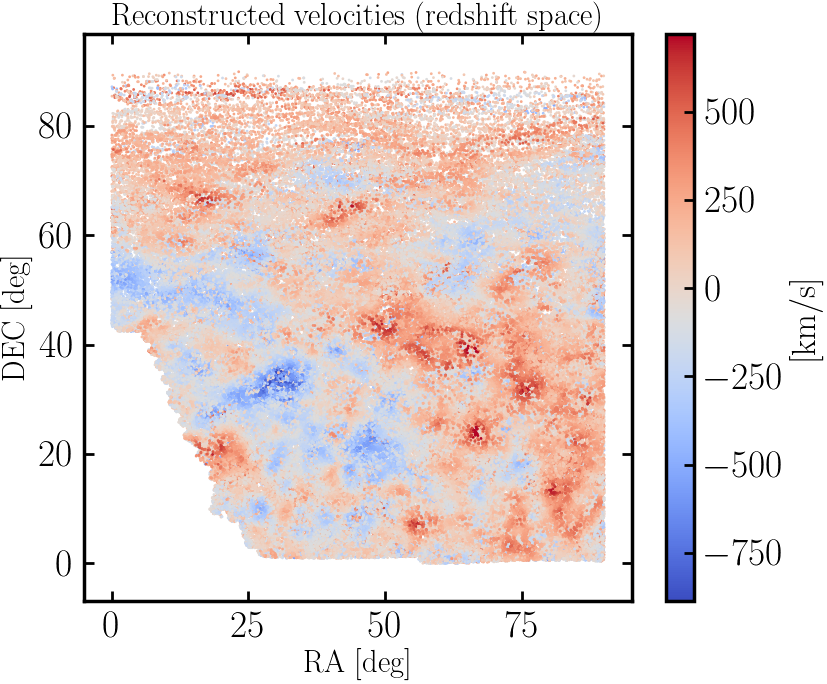}
    \includegraphics[width=0.48\textwidth]{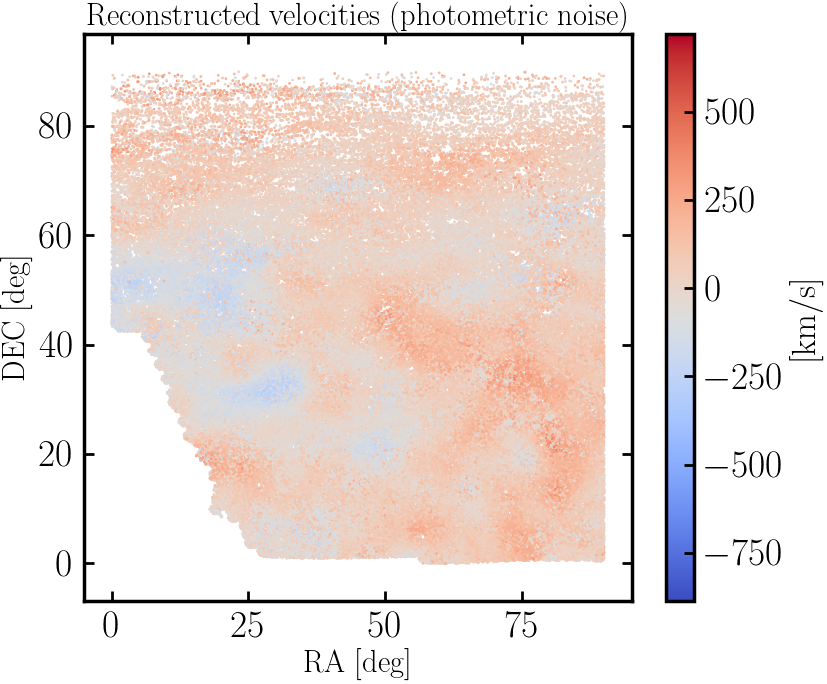} 
    \caption{Visualizations of the radial velocity reconstruction in different scenarios for the DESI NGC footprint projected onto the two-dimensional RA-DEC plane. In the top left corner, we show a map of the true velocities for galaxies with redshifts between $z = 0.505$ and $z = 0.495$. On the top right and bottom left, we show the reconstructed velocity field in real and redshift space, respectively. We see visual resemblance between truth and reconstruction on large scales in both cases, though the velocity amplitudes are lessened relative to the truth due to the smoothing applied to the maps. This effect is even more pronounced in the final case of reconstructed velocities from a noisy redshift catalog, $\sigma_z/(1+z) = 0.02$: adding noise brings the inferred galaxy overdensity field closer to white noise, reducing the available information for reconstruction. Note that when we refer to galaxy samples in redshift space, what we mean is that we have added redshift space distortions to their positions, as is the case of observed galaxies. 
    }
    \label{fig:RA_DEC_recon}
\end{figure*}

We show the comoving number densities as a function of redshift, $\bar n(z)$, for all of these tracers in Fig.~\ref{fig:nz}. 
These are extracted from  \cite{2023AJ....165..253H,2023AJ....165..126R,2023AJ....165...58Z}. 
As expected, the two LRG samples have roughly constant number density between $z = 0.4$ and $z = 0.8$, with this range being slightly expanded for the Extended LRG sample. The comoving densities are $5.2 \times 10^{-4} (\mpcoh)^{-3}$ and $1.05 \times 10^{-3} (\mpcoh)^{-3}$ for the Main and Extended catalogs, respectively. The ELG sample, on the other hand, peaks at higher redshifts, and its number densiy is roughly constant between $z = 0.85$ and $z = 1.2$, and equal to that of the Main LRG sample. We note that the number density we report for the BGS declines rapidly for $z > 0.5$. It is also lower than \cite{2023AJ....165..253H}, since here we are using a redshift distribution, $N(z)$, with additional quality and magnitude cuts that match the selection criteria for the ``BGS Bright'' sample of \cite{2023AJ....165..253H}. 

The biases of the different tracers: Main LRGs, Extended LRGs, ELGs, and BGS, as a function of reddshift are shown in Fig.~\ref{fig:bz}. The bias is computed using the \cite{2010ApJ...724..878T} empirical mass-bias relationship for the host halos. We see that the LRG bias increases slowly for $z \gtrsim 0.4$ due to the evolution of halo bias with redshift. At lower redshifts, $z \lesssim 0.5$, the bias of the LRGs seems to decrease with redshift, contrary to our expectation. As the LRG fits are performed at $z = 0.5$ and $z = 0.8$, we conjecture that this is an artifact of the extrapolation to lower redshifts rather than a property inherent to the LRG population. Relative to the Main sample, the Extended LRGs have a bias that is lower by $\Delta b = 0.1$ at $z \sim 0.5$, but overall follows the same shape. On the other hand, the BGS and ELG objects exhibit more predictable behavior with their linear bias increasing towards higher redshifts and remains significantly lower than that of the LRGs ($\bar b_{\rm LRG} \approx 2.2$, $\bar b_{\rm ELG} \approx 1.3$, $\bar b_{\rm BGS} \approx 1.5$).

Apart from adding realism to this study through modeling the galaxy-halo connection via direct fits to the data and the redshift distribution of each of the tracers, we also consider the effect of applying a survey mask.
In particular, since our main goal is to evaluate our ability to reconstruct the galaxy field in anticipation of joint DESI and ACT analysis, we adopt the public Data Release 9 (DR9) survey masks for LRGs \cite{2023AJ....165...58Z}. In particular, we construct those by coadding the \texttt{BITMASK} and \texttt{GRZMASK} fields in order to make sure that the tile has been observed once or more in each magnitude bin ($g$, $r$, $z$) with sufficient quality of the image\footnote{For more details on the DR9 masks, see:\\ \href{https://desi.lbl.gov/trac/wiki/keyprojects/y1kp3/ImagingSys/LRGMask}{https://desi.lbl.gov/trac/wiki/keyprojects/y1kp3/ImagingSys/LRGMask}}.
The full-sky mask is shown in Fig.~\ref{fig:DESI_mask} in a Mollweide projection using \texttt{HEALPix}. The two disjoint regions in the map correspond to the North Galactic Cap (NGC) and the South Galactic Cap (SGC).

We note that since the \texttt{base}-resolution light cone mocks, which make up the majority of synthetic catalogs considered in this study, cover only an octant of the sky ($0^\circ < {\rm RA} < 90^\circ$, $0^\circ < {\rm DEC} < 90^\circ$), the overlap between the DR9 mask and the octant, naively, is only about $100 \ {\rm deg}^2$. To increase the overlapping area, we rotate the galaxies in our \texttt{base} mock catalogs by $20^\circ$ in RA and $100^\circ$ in DEC, obtaining coverage of almost the entire octant, $\sim$5000 ${\rm deg}^2$. On the other hand, the \texttt{huge}-resolution light cone mocks cover the entire sky out to $z \lesssim 2.3$, and for those, we make direct use of the full DR9 mask.

\subsection{Reconstruction}
\label{sec:recon}

We can infer the reconstructed velocity field from the observed or simulated galaxy number overdensity, $\delta_g$, by solving the linearized continuity equation
in redshift space:
\begin{equation}
    \nabla \cdot \bvel + \frac{f}{b} \nabla \cdot [(\bvel \cdot \hat \bn) \hat \bn] = -a H f \frac{\delta_g}{b}
\end{equation}
where $H(z)$ is the redshift-dependent Hubble parameter, $f$ is the logarithmic growth rate, defined as $f \equiv d \ln(D)/d \ln(a)$ with $D(a)$ the growth factor and $a$ the scale factor. 
Here, we assume that the observed galaxy overdensity $\delta_g$ is related to the matter overdensity in redshift space, $\delta$, by a linear bias factor, $b$, such that $\delta_g = b \delta$. 

In order to obtain an estimate of the individual galaxy velocities, we adopt the standard reconstruction method typically utilized in BAO analysis to augment the signal around the peak \cite{ESSS07}, as per the \texttt{MultiGrid} implementation of \cite{2015MNRAS.450.3822W} via the package `pyrecon'\footnote{\url{https://github.com/cosmodesi/pyrecon}}. 
This method yields an estimate of the first-order (i.e., Zel'dovich approximation) galaxy displacement field, which we can evaluate at the location of each galaxy and convert into an estimate for the velocity. We detail the procedure below.

We can apply the standard reconstruction procedure via \texttt{pyrecon} to some tracer of the matter field by following these steps:
\begin{enumerate}
\item We smooth the tracer density field, $\delta_g$, with a Gaussian filter, $\mathcal{S}(k) =  \exp[-(kR_s)^2/2]$, of some smoothing scale $R_s$. A typical choice of the smoothing scale is $\sim$10 $\mpcoh$.
\item We then compute the 3D shifts, $\mathbf{\psi}$, by unbiasing the smoothed galaxy density field (i.e., dividing by the sum of the linear bias, $b$, and the linear redshift-space factor, $f \mu^2$) and taking the inverse gradient. In a period cubic box with a line-of-sight along the $z$ direction, this can be expressed as\footnote{Note that if the line-of-sight varies with angle, as is the case with the light cones, one needs to adopt the \texttt{MultiGrid} method}:
\begin{equation}
  \mathbf{\psi}_{\mathbf{k}} = -\frac{i\mathbf{k}}{k^2}
  \mathcal{S}(k)\ \Big( \frac{\delta_g(\mathbf{k})}{b + f\mu^2} \Big),
\label{eqn:recon_shift}
\end{equation}
where $\mu$ is the line-of-sight angle, $\mu = \hat{n} \cdot \hat{k}$. 
In the limit of very large scales, where the approximations of scale-independent bias and supercluster infall hold, the calculated shift field approaches the negative smoothed Zeldovich displacement, i.e.\  $\mathbf{\psi}_{\mathbf{k}} \approx - \mathcal{S}(k) \bPsi^{(1)}(\bk)$.
\item We next inverse-Fourier transform the shifts field, $\psi_{\mathbf{k}}$, into configuration space and evaluate it at the location of each galaxy, $\bx$.
\item Finally, we calculate the reconstructed velocity as:
\begin{equation}
  \bvel^{\rm rec}(\bx) = f(z) a(z) H(z) \psi(\bx) ,
\end{equation}
which has a component parallel and perpendicular to the line-of-sight, $\bvel = \{v_{||}, \ v_\perp\}$. For the most part, we are interested in the line-of-sight velocity, 
since that is the quantity that matters for kSZ analysis. However, we also report reconstruction statistics for $v_{\perp}$. Note that since we divide by the linear redshift-space factor, the shifts are in real space (rather than redshift space).
\end{enumerate}

In Fig.~\ref{fig:RA_DEC_recon}, we visualize the reconstructed radial velocity field. The catalog appears to be sparser for large declination angles due to projection effects. In the top left corner, we show a map of the true velocities for galaxies with redshifts between $z = 0.505$ and $z = 0.495$. While on large scales the velocities appear to be coherent, we see a lot of features on small scales due to the thermal motions of galaxies within the halos and their non-linear evolution. On the top right and bottom left, we show the reconstructed velocity field in real and redshift space, respectively. We see visual resemblance between truth and reconstruction on large scales in both cases, though the velocity amplitudes are lessened relative to the truth due to the smoothing applied to the maps. This is even more pronounced in the final case shown here, that of reconstructed velocities from a noisy redshift catalog, $\sigma_z/(1+z) = 0.02$, which brings the inferred galaxy overdensity field closer to white noise and reduces the available information for reconstruction. We note that the redshift space cases (bottom panels) appear enhanced relative to the real-space case (top right panel) due to the presence of RSD effects.

\begin{figure}[H]
    \centering
    \includegraphics[width=0.5\textwidth]{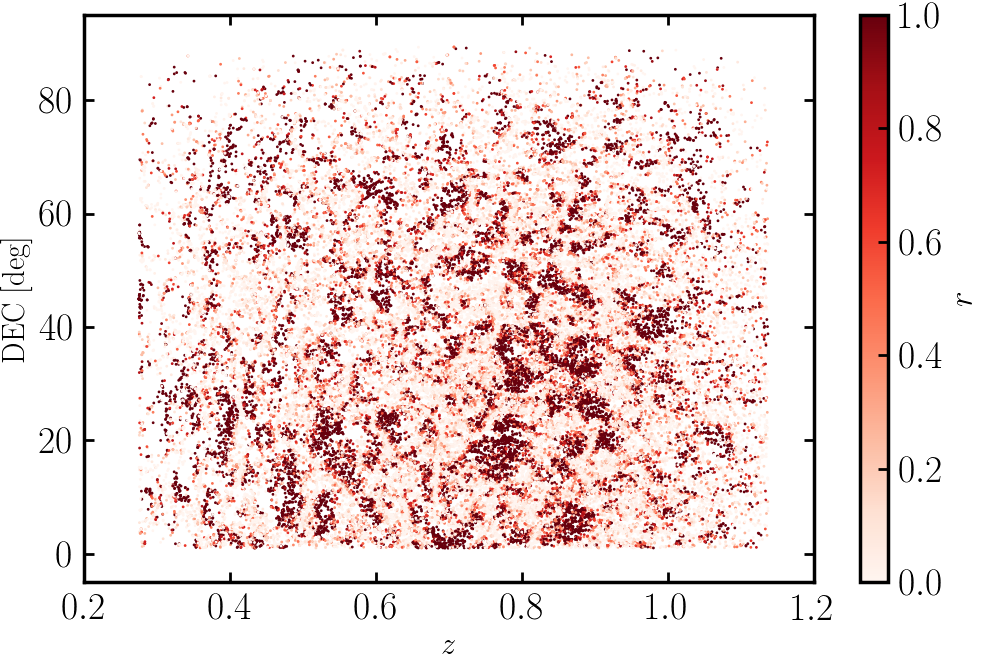}
    \caption{Visualizations of the fractional error in velocity reconstruction of a galaxy sample in redshift space without redshift errors (same as bottom left panel of Fig.~\ref{fig:RA_DEC_recon}) projected onto the two-dimensional Z-DEC plane. The fractional error is computed with respect to the true galaxy velocity as $|v^{\rm true} - v^{\rm rec}|/|v^{\rm true}|$. Similarly to Fig.~\ref{fig:RA_DEC_recon}, we also see that regions with extreme values of the velocities are reconstructed worse than regions where the velocity amplitudes are small. This is the result of the breakdown of linear theory and the smoothing applied to the galaxy overdensity. 
} 
    \label{fig:Z_DEC_recon}
\end{figure}

In Fig.~\ref{fig:Z_DEC_recon}, we visualize the fractional error in the velocity reconstruction of a galaxy sample in redshift space (same as bottom left panel of Fig.~\ref{fig:RA_DEC_recon}). The fractional error is computed with respect to the true galaxy velocity as $|v^{\rm true} - v^{\rm rec}|/|v^{\rm true}|$. We see some boundary effects near the edges of the redshift distribution corresponding to the regions where the number density of the sample is low. Similarly to Fig.~\ref{fig:RA_DEC_recon}, regions with more extreme values of the velocities are reconstructed worse than regions where the velocities are small. This is the case due to the breakdown of the linear approximation adopted in standard reconstruction and the smoothing applied to the galaxy overdensity. We also provide a linear theory analytical approximation to the decorrelation of velocities along the line-of-sight in App.~\ref{app:decorr}, finding that radial velocities for pairs of galaxies with a separation vector along the LOS decorrelate faster than radial velocities with a separation vector perpendicular to the LOS.

\subsection{The correlation coefficient}
\label{sec:corrcoeff}

An important metric for evaluating the performance of the velocity reconstruction is the correlation coefficient, $\boldsymbol{r}$, which can also be projected along and perpendicular to the line-of-sight, $\boldsymbol{r} = [r_{\perp}, \ r_{||}]$. In the case of stacked kSZ analysis, it is the component along the line-of-sight that determines the signal-to-noise of the measurement, $r_{||}$, as the sign and strength of the signal are proportional to the velocity along the line-of-sight \cite{2021PhRvD.103f3513S}. Following \cite{2021PhRvD.103f3513S}, we define the correlation coefficient as:
\begin{equation}
    r_{||,\perp} = \frac{\langle v_{||,\perp}^{\rm true} v_{||,\perp}^{\rm rec} \rangle}{\sigma_{||,\perp}^{\rm true} \sigma_{||,\perp}^{\rm rec}} , 
\end{equation}
where $\sigma_{||,\perp}$ denotes the root mean square of the velocities projected along or perpendicular to the line-of-sight. We emphasize that superscript `true' refers to the \textbf{host halo} velocities rather than the galaxy velocities. We make this choice as the kSZ signal depends on the bulk velocity of the halo rather than the individual galaxies, which may have additional thermal motions within the halo \citep[see e.g.][for a recent hydro simulation study]{2023arXiv230500992H}. 
For our fiducial case of DESI-like red galaxies at $z = 0.5$, we find that the cross-correlation coefficient between the halo and galaxy velocities is $r_{\rm gal,halo} \equiv {\langle v_{\rm 1D}^{\rm halo} v_{\rm 1D}^{\rm gal} \rangle}/({\sigma_{\rm 1D}^{\rm halo} \sigma_{\rm 1D}^{\rm gal}}) \approx 0.8$. 

Additionally, we also study the correlation coefficient as a function of scale, which holds information about the scale-dependence of the reconstruction. We can compute this quantity as follows. For a sample of galaxies with positions $\bx_i$ and line-of-sight velocities, $v_i$, we first evaluate the momentum field as:
\begin{equation}
    q(\bx) = [1 + \delta_g(\bx)] v(\bx) ,
\end{equation}
where we weight each galaxy by its velocity and employ the optimized TSC interpolation scheme from \texttt{abacusutils}\footnote{https://github.com/abacusorg/abacusutils} \citep[see Appendix of][for details]{2023arXiv230812343H}. We can then compute the power spectrum of the galaxy momentum field, $P^{q_{||}, q_{||}}$, as:
\begin{equation}
    \langle q(\bk) q(\bk') \rangle \equiv (2 \pi)^3 P^{q_{||}, q_{||}}(k) \delta_D(\bk-\bk')
\end{equation}
With this at hand, we can calculate the primary quantity of interest, the scale-dependent correlation coefficient, $r(k)$, between the true and reconstructed line-of-sight velocities:
\begin{equation}
\label{eq:r_k}
    r(k) \equiv \frac{P^{q_{||}^{\rm true}, q_{||}^{\rm rec}}(k)}{\sqrt{P^{q_{||}^{\rm true}, q_{||}^{\rm true}}(k)P^{q_{||}^{\rm rec}, q_{||}^{\rm rec}}(k)}}
\end{equation}
Since the standard reconstruction method that we adopt (see Section~\ref{sec:recon}) makes a number of approximations that hold on large scales (linear scale-independent bias, Kaiser approximation, linearized continuity equation), but break down on small and intermediate scales, we expect that the correlation coefficient between the true and reconstructed velocity will be higher on large scales.

\begin{figure}[H]
    \centering
    \includegraphics[width=0.48\textwidth]{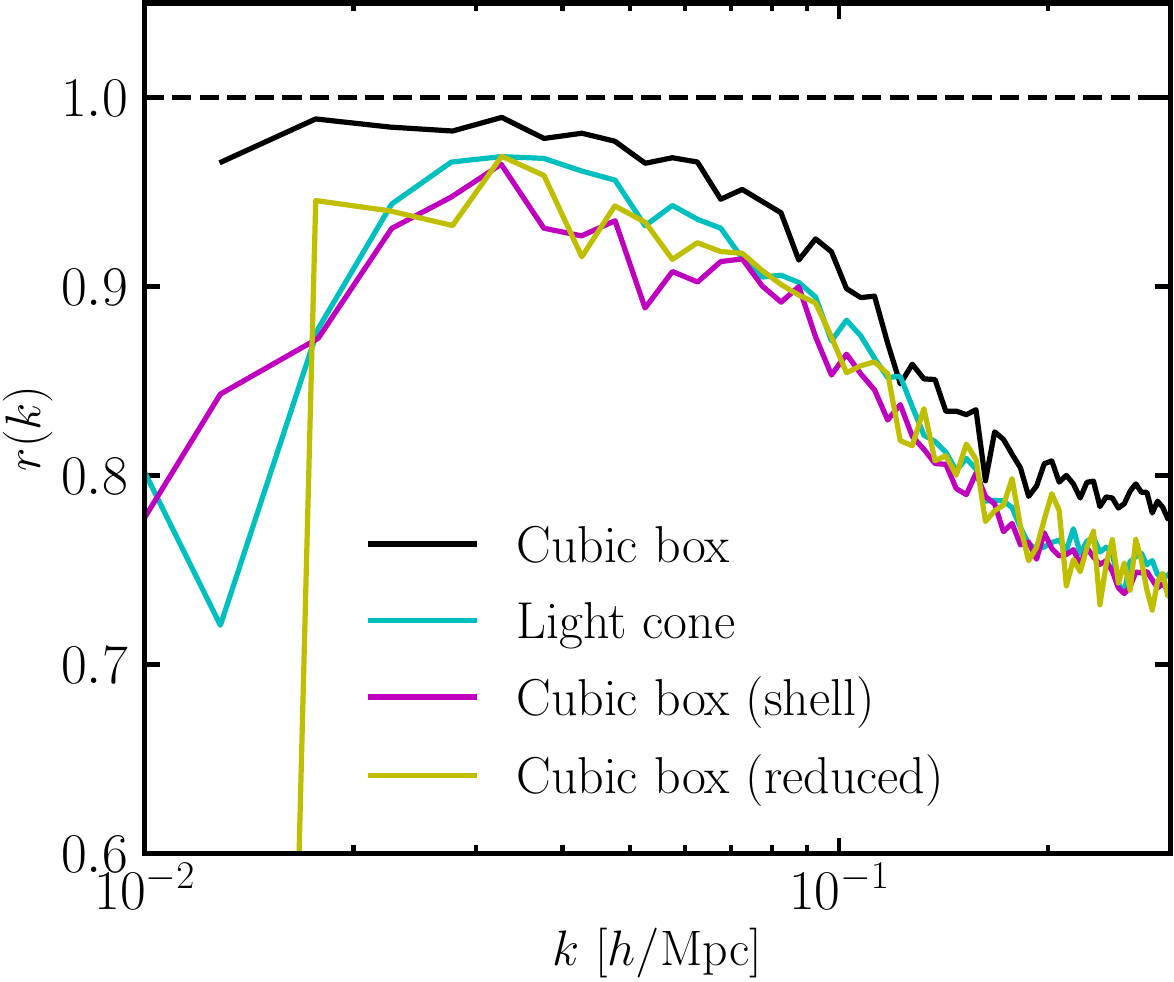}
    \caption{Cross-correlation coefficient in the line-of-sight direction as a function of scale for a galaxy sample with a light cone and a cubic box geometry (see Eq.~\ref{eq:r_k}). The galaxy sample we adopt is DESI-like LRGs at $z = 0.5$. We see that the cubic box has higher $r(k)$ on all scales compared with the other three samples. We note that this sample has constant number density and bias as well as periodic boundary conditions, which saves the need to generate randoms. It is reassuring to see the excellent agreement between the shell cutout, the reduced cubic box and the light cone. This suggests that indeed the difference between the light cone and cubic box is due to the reduction in volume (i.e., depth of the survey) and the number density variation rather than due to the redshift evolution of the sample or systematics associated with the light cone catalogs. The effect of volume reduction can be understood as a loss of large-scale information, which is crucial to reconstruction, whereas the density variation of the sample contributes shot noise and makes the density estimate more noisy, which also leads to worse reconstruction.} 
    \label{fig:r_k}
\end{figure}

In Fig.~\ref{fig:r_k}, we show the cross-correlation coefficient as a function of scale for a galaxy sample with a light cone geometry and a cubic box geometry. We study three configurations for the cubic box: one in which we apply reconstruction to the full periodic box (dubbed `Cubic box'), one in which we cut an octant shell from the cubic box with the same geometry as the light cone (dubbed `Cubic box (shell)'), and one in which we reduce the volume of the cubic box by making a cut in the $z$ (line-of-sight) direction, $0 < z < 830 \mpcoh$, which matches the volume available in the light cone catalog (dubbed `Cubic box (reduced)'). The galaxy sample we adopt is DESI-like LRGs. For the cubic box, we take the HOD catalog outputs at $z = 0.5$, whereas for the light cone, we apply an $N(z)$ downsampling that follows a Gaussian, $\mathcal{N}(0.5, \ 0.2)$. Most notable is the fact that the cubic box has higher $r(k)$ on all scales compared with the other three samples. We note that this sample has constant number density and bias as well as periodic conditions, which renders the generation of randoms obsolete. It is reassuring to see the excellent agreement between the shell cutout and the light cone. This suggests that indeed the difference between the light cone and cubic box is due to the reduction in volume (i.e., depth of the survey) and the density variation of the sample rather than due to its redshift evolution or systematics associated with the light cone catalogs. The density variation of the sample contributes shot noise and makes the density estimate more noisy, which also leads to worse reconstruction.

\section{Performance on the light cone}
\label{sec:results}

We create a number of synthetic catalogs in order to assess the performance of the reconstruction. Schematically, we divide the test scenarios into two broad categories: idealized and realistic. For the idealized tests, we employ single-tracer galaxy catalogs with simple redshift distributions (a Gaussian or a step function) with no survey mask, and study various effects in isolation pertaining to the sample properties and reconstruction parameters. For the realistic tests, we focus on studying galaxy samples with properties closely mimicking the DESI data.

\begin{table*}
\begin{center}
\begin{tabular}{ c c c c c c c c c c c }
 \hline\hline
 Test name & Tracer(s) & $N(z)$ & Area & $\frac{\sigma_z}{1+z}$ & $N_{\rm mesh}$ & $R_{\rm sm}$ & $r_\perp$ & $r_{||}$ & $r^{\rm RSD}_\perp$ & $r^{\rm RSD}_{||}$ \\ [0.5ex]
 \hline
Fiducial & Main LRG & $\mathcal{N}(0.5, 0.2)$ & Octant & $0.0$ & $1024$ & $12.5$ & \cellcolor{red!35.93} $0.72 \pm 0.013$ & \cellcolor{blue!34.77} $0.70 \pm 0.022$ & \cellcolor{red!35.94} $0.72 \pm 0.013$ & \cellcolor{blue!32.15} $0.64 \pm 0.018$ \\ [0.5ex]
Narrow $N(z)$ & Main LRG & $\mathcal{N}(0.5, 0.1)$ & Octant & $0.0$ & $1024$ & $12.5$ & \cellcolor{red!37.02} $0.74 \pm 0.014$ & \cellcolor{blue!36.82} $0.74 \pm 0.020$ & \cellcolor{red!36.89} $0.74 \pm 0.014$ & \cellcolor{blue!33.33} $0.67 \pm 0.019$ \\ [0.5ex]
ELG & ELG & $\mathcal{N}(0.8, 0.2)$ & Octant & $0.0$ & $1024$ & $12.5$ & \cellcolor{red!30.18} $0.60 \pm 0.018$ & \cellcolor{blue!28.32} $0.57 \pm 0.013$ & \cellcolor{red!31.22} $0.62 \pm 0.017$ & \cellcolor{blue!30.13} $0.60 \pm 0.012$ \\ [0.5ex]
BGS & BGS & $\mathcal{N}(0.4, 0.05)$ & Octant & $0.0$ & $1024$ & $12.5$ & \cellcolor{red!37.73} $0.76 \pm 0.022$ & \cellcolor{blue!34.71} $0.72 \pm 0.062$ & \cellcolor{red!37.45} $0.75 \pm 0.021$ & \cellcolor{blue!33.39} $0.68 \pm 0.030$ \\ [0.5ex]
High $\bar n$ & Extended LRG & $\mathcal{N}(0.5, 0.2)$ & Octant & $0.0$ & $1024$ & $12.5$ & \cellcolor{red!36.43} $0.73 \pm 0.013$ & \cellcolor{blue!34.96} $0.70 \pm 0.024$ & \cellcolor{red!36.42} $0.73 \pm 0.014$ & \cellcolor{blue!32.64} $0.65 \pm 0.018$ \\ [0.5ex]
Lowest-resolution & Main LRG & $\mathcal{N}(0.5, 0.2)$ & Octant & $0.0$ & $256$ & $12.5$ & \cellcolor{red!30.78} $0.62 \pm 0.028$ & \cellcolor{blue!23.05} $0.46 \pm 0.044$ & \cellcolor{red!30.84} $0.62 \pm 0.032$ & \cellcolor{blue!25.52} $0.51 \pm 0.038$ \\ [0.5ex]
Low-resolution & Main LRG & $\mathcal{N}(0.5, 0.2)$ & Octant & $0.0$ & $512$ & $12.5$ & \cellcolor{red!34.19} $0.68 \pm 0.017$ & \cellcolor{blue!30.84} $0.62 \pm 0.032$ & \cellcolor{red!34.24} $0.68 \pm 0.018$ & \cellcolor{blue!30.30} $0.61 \pm 0.024$ \\ [0.5ex]
High-resolution & Main LRG & $\mathcal{N}(0.5, 0.2)$ & Octant & $0.0$ & $2048$ & $12.5$ & \cellcolor{red!35.13} $0.70 \pm 0.013$ & \cellcolor{blue!34.49} $0.69 \pm 0.018$ & \cellcolor{red!34.96} $0.70 \pm 0.013$ & \cellcolor{blue!31.07} $0.62 \pm 0.018$ \\ [0.5ex]
Smoothing 1 & Main LRG & $\mathcal{N}(0.5, 0.2)$ & Octant & $0.0$ & $1024$ & $7.5$ & \cellcolor{red!37.38} $0.75 \pm 0.016$ & \cellcolor{blue!34.76} $0.70 \pm 0.026$ & \cellcolor{red!37.04} $0.74 \pm 0.017$ & \cellcolor{blue!29.75} $0.59 \pm 0.021$ \\ [0.5ex]
Smoothing 2 & Main LRG & $\mathcal{N}(0.5, 0.2)$ & Octant & $0.0$ & $1024$ & $10.0$ & \cellcolor{red!36.92} $0.74 \pm 0.013$ & \cellcolor{blue!35.44} $0.71 \pm 0.023$ & \cellcolor{red!36.80} $0.74 \pm 0.014$ & \cellcolor{blue!31.74} $0.63 \pm 0.019$ \\ [0.5ex]
Smoothing 3 & Main LRG & $\mathcal{N}(0.5, 0.2)$ & Octant & $0.0$ & $1024$ & $15.0$ & \cellcolor{red!34.84} $0.70 \pm 0.014$ & \cellcolor{blue!33.80} $0.68 \pm 0.022$ & \cellcolor{red!34.94} $0.70 \pm 0.013$ & \cellcolor{blue!31.97} $0.64 \pm 0.018$ \\ [0.5ex]
Small area & Main LRG & $\mathcal{N}(0.5, 0.2)$ & 392 deg$^2$ & $0.0$ & $1024$ & $12.5$ & \cellcolor{red!20.12} $0.40 \pm 0.080$ & \cellcolor{blue!21.28} $0.43 \pm 0.080$ & \cellcolor{red!19.64} $0.39 \pm 0.079$ & \cellcolor{blue!18.46} $0.37 \pm 0.074$ \\ [0.5ex]
Narrow step & Main LRG & $\Theta(0.4, 0.6)$ & Octant & $0.0$ & $1024$ & $12.5$ & \cellcolor{red!37.52} $0.75 \pm 0.016$ & \cellcolor{blue!36.87} $0.74 \pm 0.016$ & \cellcolor{red!37.58} $0.75 \pm 0.016$ & \cellcolor{blue!33.81} $0.68 \pm 0.019$ \\ [0.5ex]
Wide step & Main LRG & $\Theta(0.4, 0.8)$ & Octant & $0.0$ & $1024$ & $12.5$ & \cellcolor{red!37.93} $0.76 \pm 0.013$ & \cellcolor{blue!37.20} $0.74 \pm 0.012$ & \cellcolor{red!37.97} $0.76 \pm 0.012$ & \cellcolor{blue!34.07} $0.68 \pm 0.012$ \\ [0.5ex]
Photometric & Main LRG & $\mathcal{N}(0.5, 0.2)$ & Octant & $2.0$ & $1024$ & $12.5$ & \cellcolor{red!24.70} $0.49 \pm 0.017$ & \cellcolor{blue!13.12} $0.26 \pm 0.028$ & \cellcolor{red!24.81} $0.50 \pm 0.016$ & \cellcolor{blue!14.08} $0.28 \pm 0.027$ \\ [0.5ex]
Full-sky & Main LRG & $\mathcal{N}(0.5, 0.2)$ & Sphere & $0.0$ & $2048$ & $12.5$ & \cellcolor{red!36.51} $0.73 \pm 0.002$ & \cellcolor{blue!35.48} $0.71 \pm 0.000$ & \cellcolor{red!36.52} $0.73 \pm 0.002$ & \cellcolor{blue!32.53} $0.65 \pm 0.000$ \\ [0.5ex]
NGC+SGC & Main LRG & $\mathcal{N}(0.5, 0.2)$ & DESI & $0.0$ & $2048$ & $12.5$ & \cellcolor{red!35.73} $0.71 \pm 0.001$ & \cellcolor{blue!34.99} $0.70 \pm 0.000$ & \cellcolor{red!35.80} $0.72 \pm 0.001$ & \cellcolor{blue!32.10} $0.64 \pm 0.000$ \\ [0.5ex]
Evolving $b$ & Main LRG & $\Theta(0.4, 0.9)$ & Octant & $0.0$ & $1024$ & $12.5$ & \cellcolor{red!37.68} $0.76 \pm 0.011$ & \cellcolor{blue!36.62} $0.74 \pm 0.010$ & \cellcolor{red!37.84} $0.76 \pm 0.010$ & \cellcolor{blue!33.81} $0.68 \pm 0.010$ \\ [0.5ex]
Fixed $b$ & Main LRG & $\Theta(0.4, 0.9)$ & Octant & $0.0$ & $1024$ & $12.5$ & \cellcolor{red!37.84} $0.75 \pm 0.011$ & \cellcolor{blue!36.82} $0.74 \pm 0.010$ & \cellcolor{red!37.91} $0.75 \pm 0.011$ & \cellcolor{blue!33.86} $0.68 \pm 0.010$ \\ [0.5ex]
 \hline
Fiducial & Main LRG & DESI & Octant & $0.0$ & $1024$ & $12.5$ & \cellcolor{red!36.83} $0.74 \pm 0.012$ & \cellcolor{blue!35.43} $0.71 \pm 0.016$ & \cellcolor{red!36.84} $0.74 \pm 0.011$ & \cellcolor{blue!32.72} $0.65 \pm 0.014$ \\ [0.5ex]
Mask & Main LRG & DESI & DESI NGC & $0.0$ & $1024$ & $12.5$ & \cellcolor{red!36.53} $0.73 \pm 0.012$ & \cellcolor{blue!35.06} $0.70 \pm 0.016$ & \cellcolor{red!36.56} $0.73 \pm 0.012$ & \cellcolor{blue!32.43} $0.65 \pm 0.015$ \\ [0.5ex]
Photometric & Main LRG & DESI & Octant & $2.0$ & $1024$ & $12.5$ & \cellcolor{red!25.74} $0.51 \pm 0.013$ & \cellcolor{blue!14.40} $0.29 \pm 0.019$ & \cellcolor{red!25.85} $0.52 \pm 0.013$ & \cellcolor{blue!15.16} $0.30 \pm 0.018$ \\ [0.5ex]
BGS & BGS & DESI & Octant & $0.0$ & $1024$ & $12.5$ & \cellcolor{red!34.72} $0.69 \pm 0.025$ & \cellcolor{blue!32.41} $0.65 \pm 0.028$ & \cellcolor{red!34.65} $0.69 \pm 0.028$ & \cellcolor{blue!31.91} $0.64 \pm 0.030$ \\ [0.5ex]
BGS photo-z & BGS & DESI & Octant & $2.0$ & $1024$ & $12.5$ & \cellcolor{red!25.72} $0.51 \pm 0.032$ & \cellcolor{blue!9.74} $0.19 \pm 0.084$ & \cellcolor{red!25.65} $0.51 \pm 0.031$ & \cellcolor{blue!13.47} $0.27 \pm 0.067$ \\ [0.5ex]
ELG & ELG & DESI & Octant & $0.0$ & $1024$ & $12.5$ & \cellcolor{red!27.01} $0.54 \pm 0.016$ & \cellcolor{blue!25.15} $0.50 \pm 0.022$ & \cellcolor{red!28.09} $0.56 \pm 0.015$ & \cellcolor{blue!27.37} $0.55 \pm 0.018$ \\ [0.5ex]
ELG photo-z & ELG & DESI & Octant & $2.0$ & $1024$ & $12.5$ & \cellcolor{red!17.11} $0.34 \pm 0.022$ & \cellcolor{blue!7.29} $0.15 \pm 0.033$ & \cellcolor{red!17.58} $0.35 \pm 0.022$ & \cellcolor{blue!8.61} $0.17 \pm 0.033$ \\ [0.5ex]
High $\bar n$ & Extended LRG & DESI & Octant & $0.0$ & $1024$ & $12.5$ & \cellcolor{red!37.41} $0.75 \pm 0.011$ & \cellcolor{blue!35.84} $0.72 \pm 0.013$ & \cellcolor{red!37.45} $0.75 \pm 0.010$ & \cellcolor{blue!33.40} $0.67 \pm 0.011$ \\ [0.5ex]
High $\bar n$ & Extended LRG & DESI & Octant & $2.0$ & $1024$ & $12.5$ & \cellcolor{red!26.33} $0.53 \pm 0.013$ & \cellcolor{blue!14.49} $0.29 \pm 0.020$ & \cellcolor{red!26.43} $0.53 \pm 0.012$ & \cellcolor{blue!15.38} $0.31 \pm 0.019$ \\ [0.5ex]
All tracers & LRG, ELG, BGS & DESI & Octant & $0.0$ & $1024$ & $12.5$ & \cellcolor{red!37.02} $0.74 \pm 0.011$ & \cellcolor{blue!35.84} $0.72 \pm 0.012$ & \cellcolor{red!37.04} $0.74 \pm 0.010$ & \cellcolor{blue!33.22} $0.66 \pm 0.011$ \\ [0.5ex]
All tr photo-z & LRG, ELG, BGS & DESI & Octant & $2.0$ & $1024$ & $12.5$ & \cellcolor{red!25.38} $0.51 \pm 0.014$ & \cellcolor{blue!14.21} $0.28 \pm 0.022$ & \cellcolor{red!25.53} $0.51 \pm 0.013$ & \cellcolor{blue!15.01} $0.30 \pm 0.019$ \\ [0.5ex]
Evolving $b$ & Main LRG & DESI & Octant & $0.0$ & $1024$ & $12.5$ & \cellcolor{red!36.03} $0.74 \pm 0.010$ & \cellcolor{blue!34.07} $0.71 \pm 0.014$ & \cellcolor{red!36.14} $0.74 \pm 0.010$ & \cellcolor{blue!31.73} $0.65 \pm 0.012$ \\ [0.5ex]
 \hline
 \hline
\end{tabular}
\end{center}
\caption{Values of the correlation coefficient between the true and reconstructed velocities in the parallel ($r_{||}$) and perpendicular ($r_\perp$) directions for idealistic (Section \ref{sec:ideal}) and realistic scenarios (Section \ref{sec:real}). We separate these two groups of tests by a horizontal line. Additionally, we consider the effect of switching on and off RSD effects. Of most relevance to kSZ analyses using observations is the $r_{||}^{\rm rsd}$ column, as the signal-to-noise ratio is proportional to it. We discuss the results in the relevant sections.}
\label{tab:r_coeff}
\end{table*}

\subsection{Idealized mocks}
\label{sec:ideal}

While the companion paper, \cite{RiedGuachalla2023}, considers the effects of number density, bias and smoothing scale in detail and studies the dependence of the reconstruction coefficient methodically for many different scenarios, here we focus on several choices of these parameters, which are the most pertinent to the imminently planned joint analysis between DESI and ACT. Since this work focuses on light cone mocks, there are additional effects worth exploring that might evade us in a cubic box setting.

\subsubsection{Reconstruction parameters}

When applying the reconstruction algorithm of \cite{2015MNRAS.450.3822W}, there are several free parameters that need to be set, including the smoothing scale, grid resolution, and reconstruction scheme.

The grid resolution determines the number of grid points used when painting the galaxies on a mesh to compute the density fields needed for solving the continuity equation. Here, we test a handful of choices given our simulation parameters: 256, 512, 1024, and 2048. Although the differences are small, Table \ref{tab:r_coeff} suggests that the higher the resolution, the larger the correlation coefficient in both the perpendicular and parallel directions. Unsurprisingly, having a finer grid allows us to better reconstruct the smaller scales, which also contribute to the correlation coefficient (as it is derived as the ratio of power spectrum integrals across all scales). Increasing the grid size to 2048 makes marginal difference (and in fact, appears to be slightly worse than 1024), as we exhaust the information that can be extracted from small scales due to the smoothing of the field and the breakdown of the linear approximation. We, therefore, set 1024 as the default choice from hereon, which is also more optimal in terms of its computational cost.

One can also set the smoothing scale to an arbitrary value, which controls the amount of isotropic smoothing (via a Gaussian kernel) performed on the (otherwise noisy) galaxy field before reconstruction is applied (see Section~\ref{sec:recon}). One reasonable choice for eliminating this degree of freedom is to apply a Wiener filter instead of a Gaussian filter to the galaxy field such that the displacement field is optimally recovered. Such an approach has been explored in \cite{2014ApJ...788...49L,Smith2023}. However, this requires reliable modeling of the noise and clustering properties of the galaxy field, which is not always trivial. For this reason, most current BAO analyses adopt a Gaussian kernel with a smoothing scale based on empirical tests performed on galaxy mocks \citep[such is also the approach in DESI, see][]{ChenDingPaillas2023}. Nonetheless, it is possible that Wiener-filter-based reconstruction provides more optimal performance. We will explore this question in future work. 

\begin{figure}[H]
    \centering
    \includegraphics[width=0.48\textwidth]{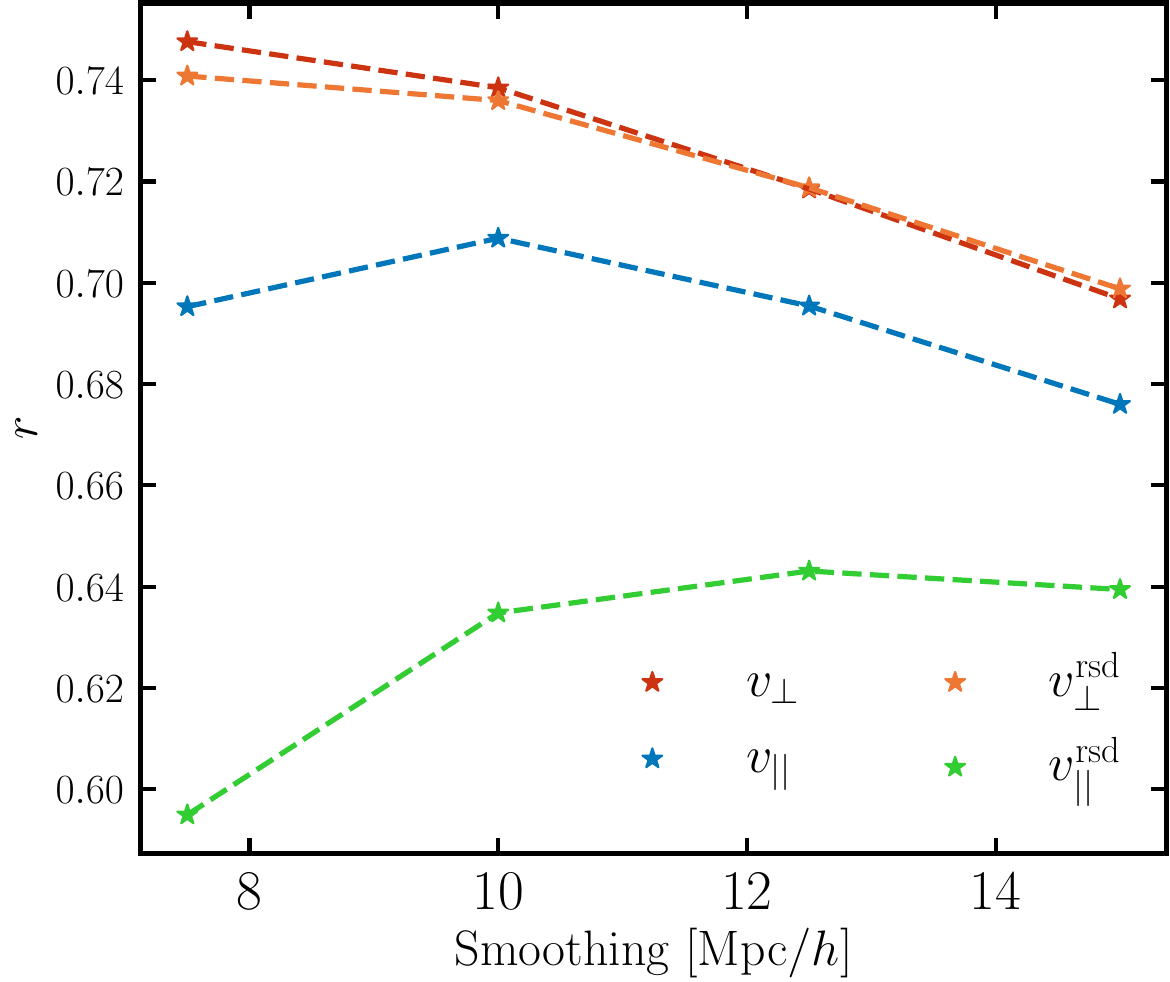}
    \caption{Correlation coefficient between the true halo velocity and the reconstructed velocity field evaluated at the galaxy locations for a LRG-like sample with a Gaussian $N(z)$ distribution, $\mathcal{N}(0.5, 0.2)$, as a function of smoothing scale. We apply Gaussian smoothing at 4 different scales: 7.5, 10, 12.5 and $15 \mpcoh$. We see that generally $r_\perp$ is higher than $r_{||}$, which we attribute to the larger variations in the density along the line-of-sight as well as on the smaller thickness. Including RSD effects leads to worse correlation coefficient along the line-of-sight, $r_{||}$, as expected. Similarly, the correlation coefficient in the perpendicular direction is largely independent of whether RSD effects are applied or not. Somewhat surprisingly, we see that while $r_\perp$ decreases as we smooth the density field and lose small-scale information, $r_{||}$ actually sees an improvement (especially when RSD effects are considered). We attribute this to the fact that the isotropic smoothing ameliorates some of the RSD effects, and leads to an improved estimate of the density field. Results are shown for the average of all 25 fiducial simulation boxes.} 
    \label{fig:smoothing}
\end{figure}

In Fig.~\ref{fig:smoothing}, we test several values of the smoothing scale: 7.5, 10, 12.5 and $15 \mpcoh$, and report the mean correlation coefficient for the 25 fiducial boxes in Table \ref{tab:r_coeff}. We note that $r$ is measured as the correlation coefficient between the true halo velocity and the reconstructed velocity field evaluated at the galaxy locations for a LRG-like sample with a Gaussian $N(z)$ distribution, $\mathcal{N}(0.5, 0.2)$. We see that generally $r_\perp$ is higher than $r_{||}$, which we attribute to the larger variations in the density along the line-of-sight as well as on the smaller thickness. Including RSD effects leads to worse correlation coefficient along the line-of-sight, $r_{||}$, as expected. Similarly, the correlation coefficient in the perpendicular direction is largely independent of whether RSD effects are applied or not. 
Naively, we expect that the higher the smoothing scale, the more small-scale information gets washed out, and thus, the lower the correlation coefficient, which is what we observe in the real-space case for both $r_{||}$ and $r_{\perp}$. Intriguingly, in redshift space, we find opposing behavior for the parallel and perpendicular correlation coefficients, $r_{||}^{\rm RSD}$ and $r_{\perp}^{\rm RSD}$, with the parallel coefficient displaying higher values with increasing smoothing scale. We speculate that isotropizing the galaxy field with the spherically symmetric Gaussian kernel reduces the RSD effects and thus leads to an improvement on the estimate of the density field and the reconstructed velocity along the line-of-sight.

Another reconstruction parameter that could potentially be varied is the reconstruction scheme used for solving the continuity equation. Modern BAO analyses typically consider the two standard schemes: \texttt{MultiGrid} (\texttt{MG}) and \texttt{IterativeFFT} (\texttt{IFFT}) \cite{2015MNRAS.450.3822W,2015MNRAS.453..456B}. We find that whether we employ the \texttt{MG} or \texttt{IFFT} scheme has negligible effect on the correlation coefficient. Thus, we opt to only show the result for the \texttt{MG} scheme. We remark that there are other approaches proposed in the literature such as smoothing and reconstructing the field iteratively, or directly solving the algebraic equation, or adopting a machine learning algorithm, or applying optimal transport \cite{2017PhRvD..96b3505S,2022MNRAS.511.1557S,2022PhRvL.129y1101N,2023MNRAS.523.6272C,2023arXiv231109531C}, which could potentially yield a non-trivial improvement on the $r$ coefficient. We leave the study of more complex reconstruction techniques for subsequent work.

\subsubsection{Number density and tracer bias}

If the galaxy field is smoothed with a Wiener filter, then how well we can reconstruct the displacement field depends on the combination $b^2 \bar{n}$. It is therefore, important to explore how this scaling translates to our quantity of interest, the velocity correlation coefficient, $r$, which receives contributions from all scales. To this end, we consider the following four samples relevant for DESI: Main LRGs, Extended LRGs, ELGs, and BGS.

Focusing on the LRGs first, we note that the Main LRG sample has a lower number density than the Extended LRG sample (see Fig.~\ref{fig:nz}), but its bias is slightly higher (see Fig.~\ref{fig:bz}). As these affects are competing with each other for these two samples, the correlation coefficient turns out to be virtually identical in both the idealized test as well as in the two realistic tests we study (see Table~\ref{tab:r_coeff}) despite the fact that the increase in number density for the high-density sample is more substantial than the decrease in bias. This implies that even if we are to greatly increase the number density, we saturate the amount of information that can be extracted from small scales due to the breakdown of the linear approximation and the smoothing of the density field. 
Nonetheless, the cross-correlation coefficient is slightly higher for the higher-density sample by about $\Delta r \approx 0.01$, including in the case where we include photometric noise (see Section~\ref{sec:real}).
On the other hand, if we substantially decrease the number density, we would see a noticeable effect on the reconstruction coefficient, as the dominant source of error would become the shot noise. For a more systematic study of the effects of bias and number density, including the effect of number density on $r$, we direct the reader to \cite{RiedGuachalla2023}.

In parallel, we explore the effect of tracer bias by studying the BGS and ELG samples. As can be seen from Fig.~\ref{fig:nz}, the BGS have a higher (though rapidly declining) number density compared with the LRGs. However, due to their relatively lower bias (see Fig.~\ref{fig:bz}), their correlation coefficients are slightly lower, as can be seen in Table~\ref{tab:r_coeff}. Intuitively, this result is expected, as the reconstruction performance scales more strongly with tracer bias than with number density. In the case of the ELGs, both the bias and the number density (as a reminder, we select the 30\% brightest ELGs) are lower, and therefore, unsurprisingly, the $r$ values are noticeably lower as well. For both of these tracers, we see interesting behavior in the line-of-sight direction: namely, $r_{||}$ is worse than $r_\perp$ and additionally, including RSD effects does not make $r_{||}$ worse as intuitively expected and seen in the LRGs case. We attribute the former effect to the larger variation of the number density along the line-of-sight and the latter effect to the smaller Finger-of-God (FoG) effects for blue galaxies \cite{2020MNRAS.499.5486A,2023MNRAS.524.2524H}. These smaller FoG effects are likely due to a combination of the dispersion velocity of the ELG and BGS being smaller (remember that velocity dispersion scales with halo mass as $M_{\rm halo} \propto v_{\rm disp}^2$), and also the fact that blue galaxies tend to be infalling (and therefore still actively star-forming), which means that their motion within the cluster is not necessarily virialized. We see a validation of this former claim, of $N(z)$ variation, when studying idealized samples for the ELG and BGS galaxies (see Table~\ref{tab:r_coeff}) that have a Gaussian $N(z)$ distribution. We find that in that case $r_{||}$ and $r_\perp$ are more consistent with each other, though we note that the line-of-sight distribution of the BGS galaxies is once again narrower and sharper than the respective ELG and LRG samples. We find evidence for the latter claim in studying reconstruction of these tracers in the cubic box and finding that when we include RSD effects, $r$ is unchanged.

\begin{figure}[H]
    \centering
    \includegraphics[width=0.48\textwidth]{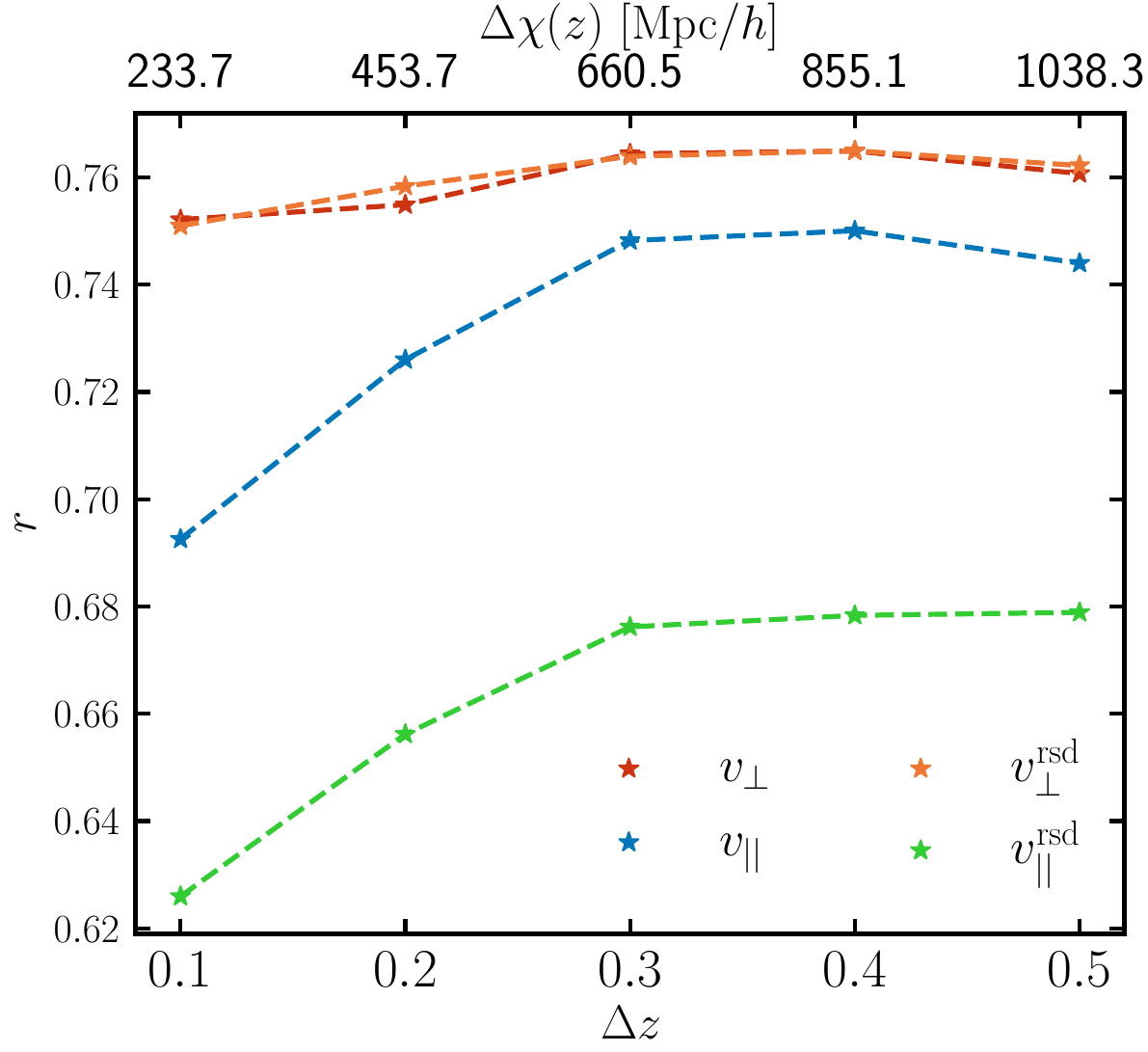}
    \caption{Correlation coefficient between the true halo velocity and the reconstructed velocity field evaluated at the galaxy locations for a LRG-like sample with a step function $N(z)$ distribution, $\Theta(z_{\rm min}, z_{\rm max})$, as a function of survey depth. Here, we set $z_{\rm min} = 0.4$ for all samples and only vary the final redshift, $z_{\rm max}$. The $x$-axis labels on the bottom show the thickness in redshift, $\Delta z$, whereas the $x$-axis labels on the top show the thickness in comoving distance, $\Delta \chi$. As before, we see that generally $r_\perp$ is higher than $r_{||}$ and that including RSD effects leads to worse correlation coefficient along the line-of-sight. Interestingly, beyond $\Delta z = 0.3$ ($\Delta \chi = 660 \mpcoh$), the correlation coefficient is largely insensitive to the survey depth, but for narrower $N(z)$ distributions, $r_{||}$, is affected by $\sim$10\%. As expected, $r_{||}$, is impacted more substantially than $r_{\perp}$, as the survey depth is measured along the line-of-sight. Results are shown for \texttt{AbacusSummit\_base\_c000\_ph002}.} 
    \label{fig:step}
\end{figure}

\subsubsection{Survey depth}

As we noted in Section~\ref{sec:corrcoeff}, the correlation coefficient drops significantly for light cone mocks compared with cubic box mocks (by $\Delta r \sim 0.1$). We argued that this is the result of the smaller volume, which reduces the amount of large-scale information (see Section~\ref{sec:recon} and references therein), and the varying number density along the line-of-sight, which leads to a poorer estimate of the density field. We test the effect of number-of-mode reduction by considering redshift distributions of varying widths, which effectively correspond to a change in the available volume in the line-of-sight direction that can be used in reconstruction.

In Fig.~\ref{fig:step}, we show the correlation coefficient between the true halo velocity and the reconstructed velocity field evaluated at the galaxy locations for a LRG-like sample with a step function $N(z)$ distribution, $\Theta(z_{\rm min}, z_{\rm max})$, as a function of survey depth. Here, we set $z_{\rm min} = 0.4$ for all samples and only vary the final redshift, $z_{\rm max}$. As before, we see that generally $r_\perp$ is higher than $r_{||}$ (due to line-of-sight density and volume variations) and that including RSD effects leads to worse correlation coefficient along the line-of-sight. Interestingly, beyond $\Delta z = 0.3$ ($\Delta \chi = 660 \mpcoh$), the correlation coefficient is largely insensitive to the survey depth, but for narrower $N(z)$ distributions, $r_{||}$, is affected noticeably, by $\sim$10\%. As expected, $r_{||}$, is impacted more substantially than $r_{\perp}$, as the density field in the line-of-sight direction is missing more large-scale modes. 


\subsubsection{Survey area}

In the vein of studying survey depth, we also consider the effect of area coverage on the correlation coefficient, $r$. We construct an area-limited sample by selecting galaxies from the fiducial LRG sample with RA and DEC between $0^\circ$ and $X^\circ$, with $X$ varying between 20$^\circ$ and 90$^\circ$ (fiducial). Naively, we expect that this should only affect the perpendicular modes and leave the parallel correlation coefficient largely unchanged. This is indeed what we observe in Table~\ref{tab:r_coeff}.

In Fig.~\ref{fig:area}, we measure the correlation coefficient between the true halo velocity and the reconstructed velocity field evaluated at the galaxy locations for a LRG-like sample with a Gaussian $N(z)$ distribution, $\mathcal{N}(0.5, 0.2)$, as a function of area coverage. The full sample covers an octant of the sky ($\sim$5200 deg$^2$, $0^\circ < {\rm RA} < 90^\circ$, $0^\circ < {\rm DEC} < 90^\circ$). We see that the correlation coefficient is largely insensitive to the area covered. Around 2000 deg$^2$, which corresponds roughly to the coverage of the DESI Y1 Galactic caps, $r$ is lower by about 3\%.

\subsection{Realistic mocks}
\label{sec:real}

In this Section, we focus on galaxy samples with properties closely mimicking the DESI data. In particular, we model LRGs, ELGs and BGS objects with a realistic DESI mask and explore effects related to their redshift noise, distribution and evolution. We also explore the response of the correlation coefficient when combining these various samples \cite{2023AJ....165..126R,2023JCAP...11..097Z}. 

\subsubsection{Redshift noise}

\begin{figure}[H]
    \centering
    \includegraphics[width=0.48\textwidth]{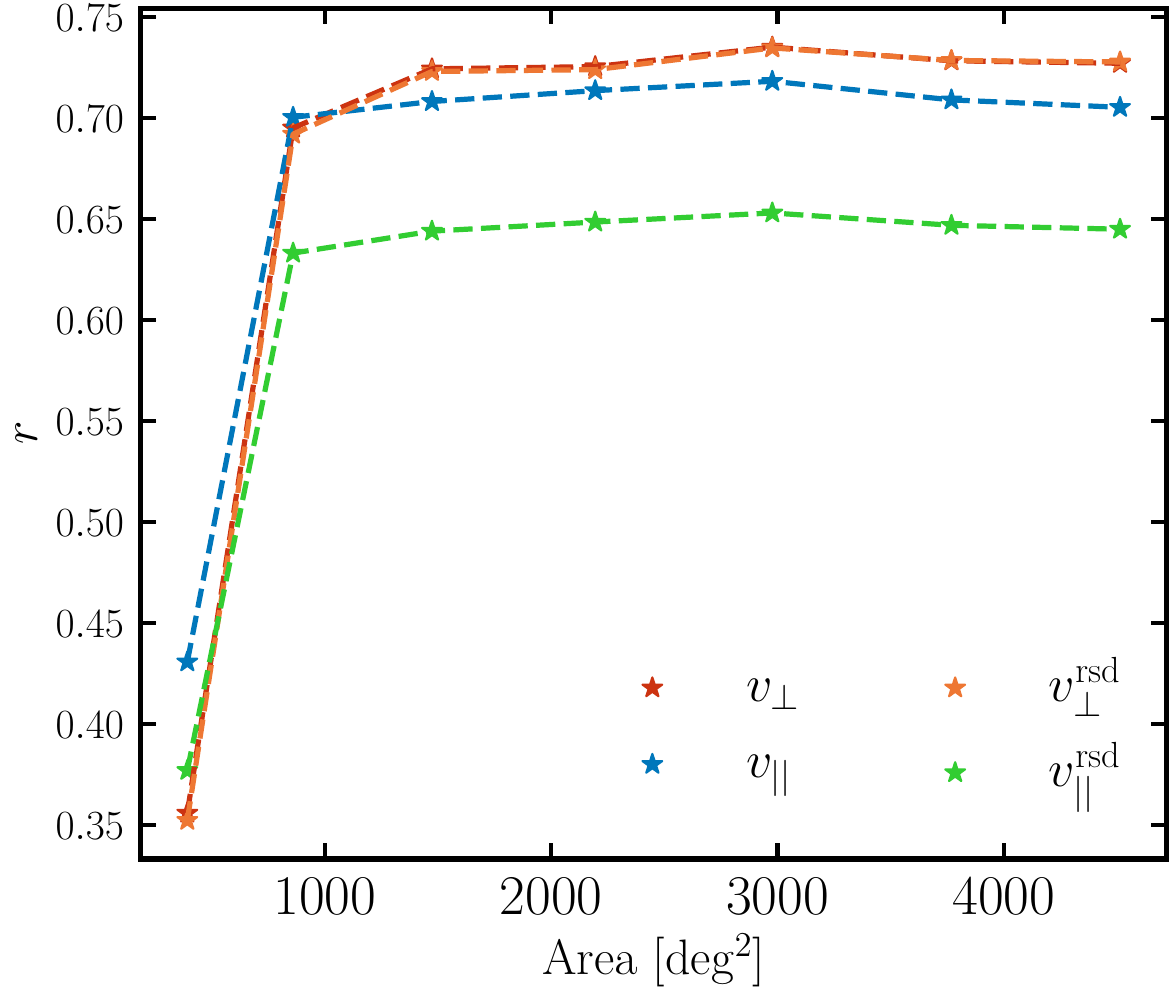}
    \caption{Correlation coefficient between the true halo velocity and the reconstructed velocity field evaluated at the galaxy locations for a LRG-like sample with a Gaussian $N(z)$ distribution, $\mathcal{N}(0.5, 0.2)$, as a function of area coverage. The full sample covers an octant of the sky ($\sim$5200 deg$^2$, $0^\circ < {\rm RA} < 90^\circ$, $0^\circ < {\rm DEC} < 90^\circ$). To create samples with different areas, we select galaxies with maximum RA and DEC of 20$^\circ$ up to 80$^\circ$. We see that the correlation coefficient is largely insensitive to the area covered. Around 2000 deg$^2$, which corresponds roughly to the coverage of the DESI Y1 Galactic caps, $r$ is lower by about 3\%. Results are shown for \texttt{AbacusSummit\_base\_c000\_ph002}.} 
    \label{fig:area}
\end{figure}

With the advancement of wide photometric surveys such as the Vera Rubin Observatory, whose redshift uncertainty goals are at the percent level, many doors will potentially open up for performing high signal-to-noise kSZ science. In addition, a number of redshift surveys, including DESI, conduct careful imaging surveys as a necessary initial step for target selection of their galaxy populations. As the redshifts derived from such surveys are based on photometry, their uncertainties are a significant inhibitor to obtaining high-fidelity cosmological constraints from photometric surveys. Therefore, a curious and important question before utilizing photometric survey data in kSZ stacking analyses is the performance of reconstruction in the line-of-sight direction. As previously quoted, we adopt constant uncertainty in the inferred photometric redshifts of all galaxy populations (note we take the 30\% of the brightest ELGs) of 2\%, i.e., $\sigma_z/(1+z) = 0.02$, which is most recently reported in \cite{2023JCAP...11..097Z}. We direct the reader to \cite{RiedGuachalla2023} for a systematic study of the effect of redshift noise for different values of $\sigma_z/(1+z)$.

Similarly to the case of survey depth, we see that adding redshift noise affects the correlation coefficient in the parallel direction significantly more than in the perpendicular direction. This is due to the smearing and isotropizing of the signal along the line-of-sight direction, which brings it closer to white noise and decreases its overall amplitude as can be seen in Fig. \ref{fig:RA_DEC_recon}. In the perpendicular direction, the correlation coefficient is not reduced to the same extent, though it is still diminished, as the full three-dimensional density field is needed to reconstruct the three-dimensional velocity field when solving the continuity equation. In particular, we see that the line-of-sight correlation coefficient is roughly reduced in half to $r_{||} \approx 0.3$, whereas in the perpendicular direction, the reduction is about 30\%, to $r_\perp \approx 0.5$.

\subsubsection{Mask effects}

So far, we have explored the idealized scenario of a survey footprint assuming no external contamination. In the case of a real survey, we usually address the issue of contamination by constructing a mask. The random sample is then generated by means of such a mask, which accounts for observational artifacts such as holes, bright stars, and incompleteness, when computing cosmology-relevant quantities. Since the focus of this study is to make forecasts for DESI, we adopt a coadded mask for DECALS DR9, consisting of a \texttt{BITWISE} mask, which flags the low-fidelity regions of the sky, as well as a mask indicating the pixels that have at least one observation per magnitude band, $grz$ (see Fig.~\ref{fig:DESI_mask}). 

In the case of the \texttt{huge} simulations, we make use of the full-sky DESI maps (i.e., both NGC and SGC), since we have full-sky light cones for those. However, the \texttt{base} simulations light cones provide coverage for only an octant of the sky, and thus, we rotate them to fit the NGC footprint and successfully fit 95\% of it. We then apply the coadded mask to both the galaxies and the randoms by querying each object to determine whether to keep it or throw it out of the sample. Utilizing the masked randoms during reconstruction ensures that the galaxy density field of the masked galaxies remains unbiased.

When performing reconstruction on the masked \texttt{base} catalogs, we find, somewhat surprisingly, that the correlation coefficient is lowered by only a small amount ($\Delta r \approx 0.01$). We note that this holds true both for the \texttt{huge} simulations, for which we apply the full DESI footprint, as well as for the \texttt{base} simulations, which utilize only the Northern Galactic Cap. This suggests that boundary effects and near-mask effects are mostly negligible and reasonably dealt with in the standard reconstruction algorithm.

\subsubsection{Bias evolution}
\label{sec:bias}

An advantage of the light cones, which cannot be faithfully reproduced in the cubic box is that they allow us to model the effect of redshift evolution of the galaxy population. In solving the continuity equation for reconstruction, there are several quantities that evolve with redshift: the growth rate, $f$, the scale factor, $a$, the Hubble parameter, $H$, and the galaxy bias, $b$. The standard assumption is that these quantities vary slowly with redshift and therefore, can be fixed to their respective effective values, typically estimated at the mean redshift of the survey. If the redshift distribution is fairly narrow (as has been the case for most spectroscopic surveys until now), one can safely neglect the redshift dependence of these parameters. However, next-generation experiments such as DESI will collect a large number of spectroscopic redshifts across a wide redshift range, which would strain the assumption of redshift insensitivity. Since reconstruction in the context of the BAO analysis is a standard tool for current redshift surveys, significant amount of effort is dedicated to studying the sensitivity of BAO constraints on various physical and observational effects including redshift evolution. For example, \cite{ChenDingPaillas2023} find that for realistic DESI-like samples the effect of structure growth evolution on the BAO parameters is tiny. However, to the best of our knowledge, the effect of bias evolution on reconstruction has not been explored in detail yet via realistic redshift-evolving light cones and a redshift-dependent modeling of the galaxy population. For a reminder on our galaxy-halo assignment procedure, see Section~\ref{sec:galaxy}.

While in the rest of the paper, we assume that the biases are fixed ($b_{\rm LRG} = 2.2$, $b_{\rm ELG} = 1.3$), in this section, we follow a slightly more complex procedure. Adopting the \cite{2010ApJ...724..878T} empirical relation between halo mass and bias, we calculate the bias of the BGS, ELG and LRG halo hosts as a function of redshift (since we save the \textsc{CompaSO} host halo mass of each galaxy). We show the bias evolution of the three samples in Fig.~\ref{fig:bz}. The bias of the LRGs, which are the most relevant tracer for kSZ analysis, ranges between $b \approx 2$ and 2.5, with an overall trend of increase with redshift, as expected. We note that below $z \lesssim 0.4$, the bias exhibits a slightly unusual behavior (it increases), which we attribute to the fact that we are extrapolating the fitted HOD parameters below $z < 0.5$. Additionally, the biases of the Extended LRG, ELG and BGS samples are lower than the Main LRG sample, as they preferentially occupy halos of smaller mass, and thus bias. To obtain the bias field in the curved sky setting, we first paint a mesh giving each galaxy a weight equal to $b(z)$ (interpolated from the curves in Fig.~\ref{fig:nz}), yielding a field proportional to $(1 + \delta_g(\bx)) b(\bx)$.  We can then obtain the bias field as the ratio between the weighted and the unweighted fields.
Thus, the matter field can be estimated as $\delta_m(\bx) = \delta_g(\bx)/b(\bx)$, which we use when solving the continuity equation. 

We report the result in Table \ref{tab:r_coeff}. In the case of the LRGs, we see that incorporating the redshift dependence of the bias leads to a very small and sometimes even unnoticeable improvement of only $\Delta r \sim 0.01$. 
There are several caveats to note in this test. Since halo mass is not an observable, when estimating the bias of the real data, we would need to take a different approach. Most commonly, one splits the galaxies into thin redshift bins and measures the ratio of the galaxy power spectrum to the predicted matter power spectrum at some benchmark cosmology. Note that our galaxy selection may not reflect the true bias of ELGs found in observations, since we interpolate the HOD parameters from only two pivot redshifts, and the true bias might evolve more dynamically.
In addition, we incorporate the redshift dependence of the pre-factor, $f(z) a(z) H(z)$, when evaluating the reconstructed velocity of each galaxy. We find that this makes a negligible difference to the final quanity of interest $r$. Note, however, that in solving the continuity equation, we assume a constant $f$. We leave a more sophisticated study, which treats the redshift dependence of the growth factor, bias and cosmological parameters natively, for the future.

\subsubsection{Combining different tracers}

Intuitively, we expect that a sample with a high number density should yield better reconstruction and thus better $r$, as it would be less affected by shot noise. One can then conjecture that including all galaxy tracers available in a survey -- e.g., LRGs, ELGs, and BGS in the case of DESI, should potentially improve the reconstructed velocities. 
Motivated by this, we perform a test in which we combine all three galaxy samples with their respective $N(z)$ distributions into a single sample with a corresponding random sample following the joint $N(z)$. While the reconstruction of the displacement field is computed using the combined set of tracers, we interpolate the displacement field and therefore the reconstructed velocity field solely at the galaxy positions of the LRGs, as this is the main sample of interest for kSZ analysis. Note that on their own, both the BGS and the ELGs perform worse than the LRGs alone, but including their three-dimensional positions ought to improve the density estimate. As can be seen in Table~\ref{tab:r_coeff}, 
the combined sample performs marginally better in the parallel direction, $\Delta r_{||} \approx 0.01$, but a) this is not a substantial increase (i.e., it is within the noise across the 25 realizations) and b) it comes at the price of a significantly more complex modeling of the bias and redshift evolution of the sample, which renders this exercise hardly worthwhile.


\section{Summary}
\label{sec:conc}

Extracting the kSZ signal around clusters through joint CMB and LSS studies will provide one of the cleanest probes of baryonic feedback, which is crucial not only for understanding gas thermodynamics and the processes that shape galaxy formation and evolution, but also for performing high-precision cosmological inference from measurements affected by the distribution of baryons (e.g., weak lensing). Since the goal of stacked kSZ analysis is to infer the baryon profiles around massive halos and since the kSZ signal is proportional to both the optical depth and the radial velocity, one can independently estimate the velocity field from the large-scale distribution of galaxies via some ``reconstruction'' method and thus eliminate the dependence on the velocity. Naturally, the success of the elimination depends on how accurately one has reconstructed the velocity field. 

In this work, we adopt the standard technique used in BAO analysis for reconstructing the Zeldovich displacements (and velocities) of galaxies through the linearized continuity equation and assess its performance in a number of realistic scenarios relevant to currently planned joint DESI and ACT analyses. In particular, we employ synthetic high-accuracy galaxy catalogs on the light cone, representative of DESI targets, and investigate the effect on reconstruction of variations in the number density, bias, mask, area, redshift noise, and survey depth, as well as modifications to the settings of the standard reconstruction algorithm. We summarize our main findings below.

\begin{itemize}
    \item When comparing the cross-correlation coefficient obtained from the light cone catalogs with the cubic box (see Fig.~\ref{fig:r_k}), we find that $r(k)$ is noticeably lower on large scales for the light cone compared with the periodic box. We attribute this to the more limited volume of the light cone, which is confirmed by the agreement with the results from the cubic box cutouts, and the variation in the number density, $N(z)$, along the line-of-sight. This latter effect implies that the shot noise term becomes more important in low-density regions, so reconstruction performs poorly.
    \item We study how well we can reconstruct the velocity field for three different tracers: LRGs, BGS and ELGs, which have different number densities and biases (see Fig:~\ref{fig:nz} and Fig:~\ref{fig:bz}). We find that the ELGs for which both the number density and the bias are lower perform significantly worse than the other two tracers, whereas the BGS have slightly lower $r$ coefficient than the LRGs despite having a much higher number density. We comment that it is the bias rather than the number density, that affects reconstruction the most. As studied in detail in the companion paper of \cite{RiedGuachalla2023}, the rule-of-thumb that reconstruction depends on the combination $b^2 \bar n$ no longer holds, as past a certain density threshold, we are dominated by the breakdown of the Zel'dovich approximation. The higher-density LRG sample, which we dub ``Extended LRGs'' does indeed perform marginally better, but the improvement in $r$ is rather modest, $\Delta r \approx 0.01$. These results are summarized in Table~\ref{tab:r_coeff}.
    \item In Fig.~\ref{fig:smoothing}, Fig.~\ref{fig:step}, and Fig.~\ref{fig:area},
    we explore the effect on the correlation coefficient for three different settings: 1) Smoothing scale: Fig.~\ref{fig:smoothing} indicates that when increasing the smoothing scale used in reconstruction, the $r$ coefficient decreases in most cases except for when we include RSD effects and measure it in the parallel direction. We conjecture that standard reconstruction yields a better result due to the isotropizing effect that the Gaussian smoothing has on the density field. Otherwise, $r$ decreases with increasing smoothing scale as we lose increasingly more small-scale information. 2) Survey depth: In Fig.~\ref{fig:step}, we study how strongly $r$ depends on the survey depth. We find that past a certain range, roughly $\Delta z \gtrsim 0.3$, having a deeper survey makes no difference to the reconstruction. One ought to be more cautious when considering shallower surveys, as the line-of-sight correlation coefficient is affected noticeably. 3) Survey area: In~\ref{fig:area}, we study the dependence of reconstruction on the area of the survey, finding that as long as our survey covers at least $\sim$1000 deg$^2$ or more, we are not very sensitive to sky coverage. There is a steady increase between $\sim$1000 and 3000 deg$^2$ after which $r$ saturates.
    \item In Section~\ref{sec:real}, we consider several effects that might affect reconstruction and thus the cross-correlation parameter $r$, including the mask of the survey, the redshift evolution of the bias and redshift noise. Intriguingly, both the bias evolution and the survey mask have virtually completely negligible effects on the reconstruction, but incorporating redshift noise as $\sigma_z/(1+z) \approx 0.02$, which reflects the latest estimate of the error budget on the photometric DESI target selection LRG sample, halves the value of $r$ in the line-of-sight direction and reduces the value of $r$ in the perpendicular direction by 30\%. In addition, we test how throwing in all three galaxy samples affects reconstruction and find that such a choice is hardly worthwhile, as it significantly increases the complexity of the model while not giving a substantial benefit to the $r$ value.
\end{itemize}


Our hope is that 
the results reported in this study will provide guidance for the optimal performance of velocity reconstruction in anticipation of upcoming joint survey analyses between CMB and LSS experiments.

\section*{Acknowledgments}

We thank Martin White and Arnaud de Mattia for making their BAO reconstruction code, \texttt{pyrecon}, publicly available. We Thank Kendrick Smith, David Valcin and Hee-Jong Seo for useful discussions about velocity reconstruction from the galaxy number density field. We thank Rongpu Zhou for useful discussions about the DESI LRG samples and related photo$-z$ errors.
B.H. thanks the Miller Institute for financially supporting her postdoctoral research.
S.F. is supported by Lawrence Berkeley National Laboratory and the Director, Office of Science, Office of High Energy Physics of the U.S. Department of Energy under Contract No.\ DE-AC02-05CH11231.
B.R. and E.S. received support from the U.S. Department of Energy under contract number DE-AC02-76SF00515 to SLAC National Accelerator Laboratory. This research used resources of the National Energy Research Scientific Computing Center (NERSC), a U.S. Department of Energy Office of Science User Facility located at Lawrence Berkeley National Laboratory, operated under Contract No. DE-AC02-05CH11231.

\bibliographystyle{prsty.bst}
\bibliography{main}

\appendix

\newpage

\section{Decorrelation of the linear velocity field}
\label{app:decorr}
To build intuition about the properties of the linear velocity field, we compute the velocity correlation function in linear theory and in the flat-sky approximation.

The cosmic \textit{peculiar} velocity is defined as the conformal time derivative of the comoving position $\bx$ related to the matter, i.e. it represents the physical velocity once the average expansion has been subtracted:
\be
\boldv \equiv \frac{d \bx}{d \eta} = \frac{d \boldr}{dt} - H(z) \boldr
\ee
If in Fourier space we define $\boldv$ such that $\boldv = \hat{k} v$, with power spectrum
\be
P_{vv} (k, z) = \left( \frac{a H f}{k} \right)^2 P^{\rm lin}_{\delta \delta}(k, z)
\ee
It can be shown that this is a very good approximation even in the mildly non-linear regime, up to $k \sim 0.5 h$/Mpc.
Then we can compute the correlation function of $\boldv$:
\ba
& \langle & v_\alpha(0) v_\beta(\boldr) \rangle 
  = \int \frac{d^3\bk}{(2\pi)^3} \hk_\alpha \hk_\beta P_{vv}(k) e^{i \bk \cdot \boldr} \\
  & = &\int \frac{d^3\bk}{(2\pi)^3} \hk_\alpha \hk_\beta P_{vv}(k) \sum_{L \geq 0} i^L (2L+1) j_L(kr) P_L({\hat k} \cdot {\hat r}) \nonumber
\ea
using Rayleigh expansion. The angular integral can be done as follows:
\be
\int d\Omega_{\hk} \hk_\alpha \hk_\beta P_L(\hk\cdot\hr) = \left\{ \begin{array}{cl}
  \frac{4\pi}{3} \delta_{\alpha\beta} & \mbox{if $L=0$} \\
  \frac{4\pi}{15} \left( 3 \hr_\alpha \hr_\beta - \delta_{\alpha\beta} \right) & \mbox{if $L=2$} \\
  0 & \mbox{if $L\ne 0,2$ }
\end{array} \right.
\ee
This gives:
\begin{widetext}
\ba
\langle v_\alpha(0) v_\beta(\boldr) \rangle 
  &=& \int \frac{k^2 dk}{(2\pi)^3} P_{vv}(k) \left( \frac{4\pi \delta_{\alpha\beta}}{3} j_0(kr) 
        - \frac{4\pi (3 \hr_\alpha \hr_\beta - \delta_{\alpha\beta})}{3} j_2(kr) \right) \nonumber \\
  &=& \psi_0(r) \delta_{\alpha\beta} + \psi_2(r) \left( \frac{3}{2} \hat r_\alpha \hat r_\beta - \frac{1}{2} \delta_{\alpha\beta} \right)
\ea
\end{widetext}
where we have defined
\ba
\psi_0(r) &=& \frac{1}{3} \int \frac{k^2 dk}{2\pi^2} P_{vv}(k) j_0(kr) \\
\psi_2(r) &=& -\frac{2}{3} \int \frac{k^2 dk}{2\pi^2} P_{vv}(k) j_2(kr)
\ea
Note that $3 \psi_0(r)$ is the usual velocity correlation function (Fourier transform of the power spectrum). 
Indeed, this is recovered with 
\be
\langle \boldv(0) \cdot \boldv(\boldr) \rangle = \sum_\alpha \langle v_\alpha(0) v_\alpha(\boldr) \rangle = 3 \psi_0(r)
\ee

\begin{figure}[H]
\centerline{\includegraphics[width=9.5cm]{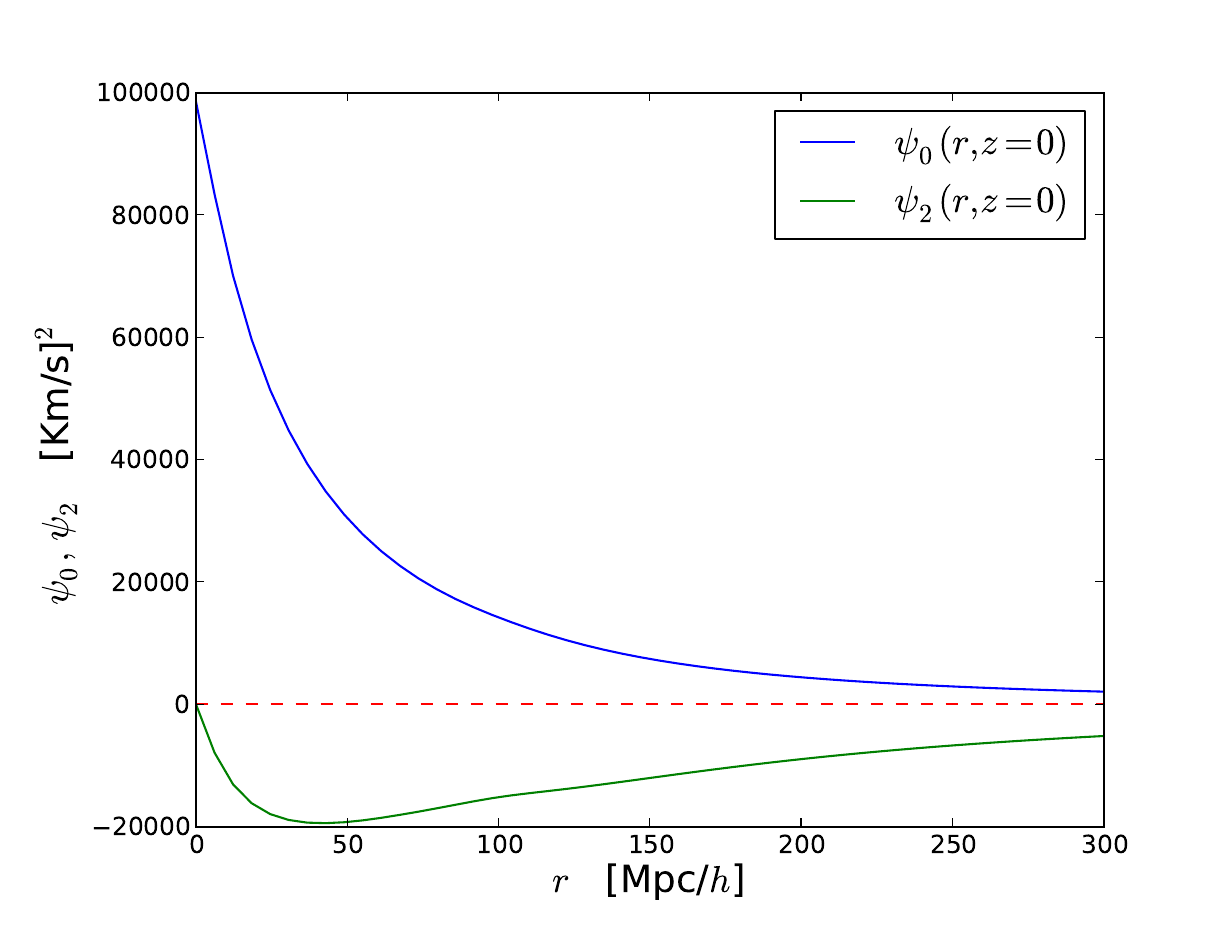}}
\caption{Real space velocity correlation functions $\psi_0(r)$ and $\psi_2(r)$ at $z = 0$.}
\label{fig:psi02}
\end{figure}

Specializing to the correlation of \textit{radial} velocities (i.e. along the line of sight $\hat{n}$) and neglecting the sky curvature:
\be
\langle v_{\hat{n}}(0) v_{\hat{n}}(\boldr) \rangle = \psi_0(r) + \psi_2(r) \left(\frac{3}{2} (\hat{n} \cdot \hat{r})^2 - \frac{1}{2} \right)
\ee
Note that this correlation does \textit{not} only depend on the separation $r$, but also on the direction of the separation vector, and in particular on its projection along the line of sight. The advantage of this form is that the function $\psi_0(r)$ and $\psi_2(r)$ can be precomputed and this makes obtaining a numerical estimate of the correlation function very fast.

For example, we can compute the correlation function of $v_r$ for points whose separation lies parallel or perpendicular to the line of sight (los):
\ba
\langle v_{\hat{n}}(0) v_{\hat{n}}(\boldr) \rangle_{\boldr \perp {\rm los}} &=& \psi_0(r) - \frac{1}{2} \psi_2(r) \\
\langle v_{\hat{n}}(0) v_{\hat{n}}(\boldr) \rangle_{\boldr \parallel {\rm los}} &=& \psi_0(r) + \psi_2(r) 
\ea

\begin{figure}
\centerline{\includegraphics[width=9.5cm]{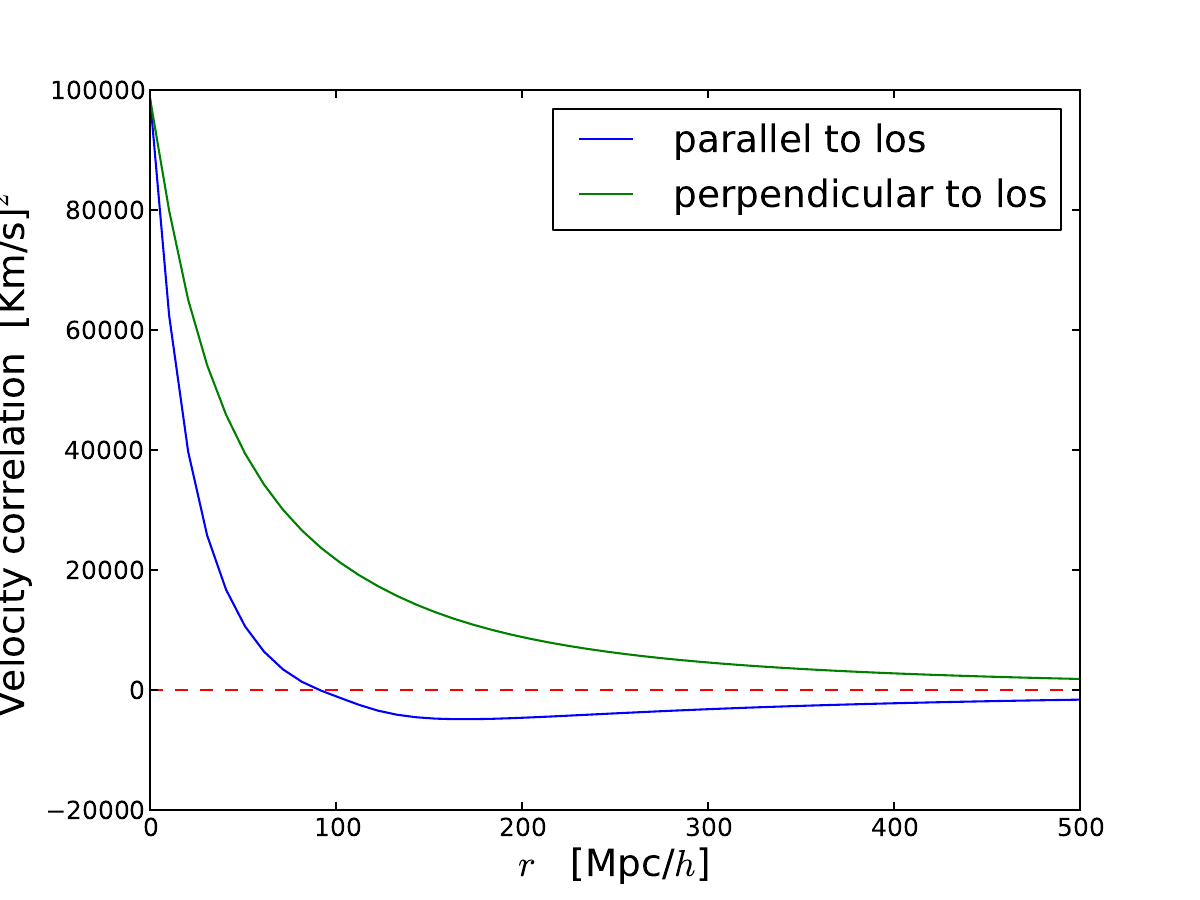}}
\caption{Real space \textit{radial} velocity correlation functions for points with separation parallel and perpendicular to the los. As we can see, radial peculiar velocities decorrelate faster than transverse velocities, which has important implications as discussed in the main body of the paper.}
\label{fig:corr}
\end{figure}

\newpage

\end{document}